\documentclass[12pt]{article}
\usepackage{a4wide}

\usepackage[dvipdfmx,hiresbb]{graphicx} 
\usepackage{mediabb}                    

\usepackage{amsmath,amssymb}
\usepackage{color}
\usepackage[bookmarks,bookmarksnumbered]{hyperref}
\usepackage{cellspace}
\usepackage{tikz}
\usepackage{mathtools}
\usepackage{cite}
\usetikzlibrary{fadings}
\usetikzlibrary{patterns}
\usetikzlibrary{shadows.blur}
\usetikzlibrary{shapes}
\usepackage{ascmac}
\usepackage{mathrsfs}
\usepackage{float}
\usepackage{adjustbox}

\newcommand{\aln}[1]{\begin{align}#1\end{align}}
\newcommand{\nn}{\nonumber\\}

\begin{document}
\title{\vspace{-3cm}
\vbox{
\baselineskip 14pt
\hfill \hbox{\normalsize 
}}  \vskip 1cm 
\bf \Large Landau theory for lattice higher gauge theory and Kramers-Wannier   duality
\vskip 0.5cm
}
\author{
Kiyoharu Kawana\thanks{E-mail: \tt kkiyoharu@kias.re.kr}
\bigskip\\
\normalsize
\it 
School of Physics, Korean Institute for Advanced Study, Seoul 02455, Korea
\smallskip
}
\date{\today}

\maketitle   
\begin{abstract} 
We derive a Landau field theory for a lattice higher gauge theory defined on $p$-dimensional open cells~(i.e., sites, links, faces, cubes, etc.), and study its continuum-limit and phases.  
In this approach, the $p$-dimensional Wilson-surface operator of the higher gauge theory is promoted to a fundamental functional field that is charged under the $p$-form global symmetry.   
By explicitly solving the functional equation of motion, we show that the classical solution exhibits the area~(perimeter) law in the strong (weak) gauge coupling limit. 
In the deconfined phase, we also construct topological defects for both $\mathrm{U}(1)$ and $\mathbb{Z}_N^{}$ higher gauge theories, as analogs of vortex and domain-wall solutions in conventional field theories with $0$-form global symmetries.    
%
Besides, we examine low-energy effective theory by identifying the phase modulations of the functional field as low-energy modes, and discuss the Coleman-Mermin-Wagner theorem for higher-form global symmetries.      
Finally, we discuss infrared duality among Landau field theories, which originates from Kramers-Wannier duality in lattice higher gauge theories.  
%

\end{abstract} 

\setcounter{page}{1} 

\newpage  

\tableofcontents   

\newpage  

\section{Introduction}\label{Sec:intro}

In modern physics, the concept of symmetry has been extended and generalized in various  directions~\cite{Gaiotto:2014kfa,Kapustin:2005py,Pantev:2005zs,Nussinov:2009zz,Banks:2010zn,Kapustin:2013uxa,Aharony:2013hda,Kapustin:2014gua,Gaiotto:2017yup,
McGreevy:2022oyu,Brennan:2023mmt,Bhardwaj:2023kri,Luo:2023ive,Gomes:2023ahz,Shao:2023gho,Hayashi:2024yjc,Hidaka:2024kfx}.  
The awareness of these generalized symmetries has broadened our understanding of quantum phases of matters and motivated us to explore new paradigm beyond conventional Landau theory for particles, i.e., field theory with  $0$-form global symmetries.     

In this paper, we derive a Landau field theory for a lattice higher gauge theory 
defined on $p$-dimensional open cells in a $D$-dimensional Euclidean hypercubic lattice. 
Here, $p$-dimensional open cells include sites~($p=0$), links~($p=1$), faces~($p=2$), cubes~($p=3$), and their higher-dimensional ones.  
Hereafter, we refer to $p$-dimensional open cell as {\it $p$-link} as in ordinary lattice gauge theories.  
The higher gauge theory defined on $p$-links possesses a $p$-form global symmetry, whose order parameter is the vacuum expectation value of the $p$-dimensional Wilson-surface operator $W[C_p^{}]$. 
By performing the Hubbard-Stratonovich~(HS) transformation to the higher gauge theory, we derive a scalar field theory defined on the space of $p$-dimensional (self-avoiding) closed surfaces.  
%
This is an extension of the previous works~\cite{Yoneya:1980bw,Kawana:2024qmz}, where the Landau theory of $\mathrm{SU}(N)$ (lattice) gauge theory was formulated in the same manner.   
%

%
We then examine the classical continuum-limit of the Landau theory and study  phases of the higher gauge theory at the mean-field level.  
One of the crucial findings is that the kinetic term of the scalar functional takes the form of a d'Alembert operator constructed from the so-called area derivatives~\cite{Migdal:1983qrz,Makeenko:1980vm,Polyakov:1980ca,Kawai:1980qq,Rey:1989ti,Iqbal:2021rkn,Hidaka:2023gwh,Kawana:2024fsn}. 
%
%
%
By solving the functional equation of motion, we show that the classical solution in the Landau theory exhibits the area (perimeter) law in the strong~(weak) gauge coupling limit, corresponding to the confined (deconfined) phase of the original higher gauge theory, as expected. 
In the broken or deconfined phase, the phase modulations in the scalar field are identified as low-energy fluctuation modes, and we examine their low-energy effective theory. 
In the case of compact $\mathrm{U}(1)$ $p$-form gauge theory, a naive effective theory is the $p$-form Maxwell theory describing a gapless $p$-form field $A_p^{}$, but the existence of topological defects (or monopoles for $D=p+2$) can render these fluctuation modes massive and  prevent the spontaneous breaking of the $p$-form global symmetry for $p\geq D-2$. 
This is the Coleman-Mermin-Wagner theorem for compact $\mathrm{U}(1)$ higher gauge theories~\cite{Gaiotto:2014kfa,Lake:2018dqm}. 
In the case of finite $p$-form gauge theory, on the other hand, the low-energy effective theory is found to be a $\mathrm{BF}$-type topological field theory  coupled to topological defects, exhibiting  topological order in general.  
%
%
%

Following the mean-field analysis, we further discuss Kramers-Wannier (KW)~duality~\cite{PhysRev.60.252,Kramers:1941zz,Itzykson:1989sx,Kaidi:2021xfk,Koide:2021zxj} in finite higher gauge theories and its implication for infrared (IR)  duality in continuum Landau field theories.   
%
In general, the KW duality asserts that the higher gauge theory with an abelian group $G$ defined on $p$-links is dual (or isomorphic) to the ``{\it gauged}" higher gauge theory defined on $(D-p-2)$-links with the Pontrjagin dual $\widehat{G}\simeq G$ as  its gauge group.  
For example, for $D=2$ and $p=0$, it corresponds to the famous KW duality between the $2$-dimensional $\mathbb{Z}_2^{}$ Ising model and the $2$-dimensional gauged $\mathbb{Z}_2^{}$ Ising model. 
We derive this KW duality in the present lattice formulation and elucidate how it can lead to an IR duality between Landau field theories in the continuum-limit.         
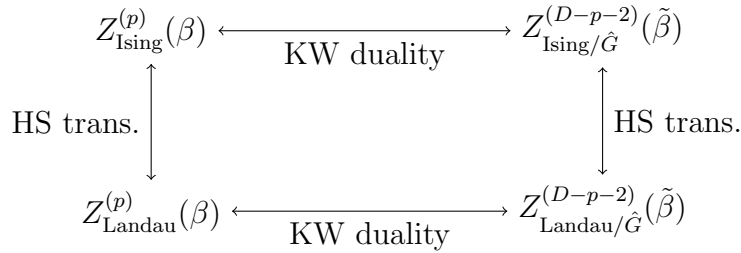
\begin{figure}
\centering 
\begin{tikzpicture}[scale=1.2,baseline=(current bounding box.center)]
  \node (A) at (0,0) {$Z_{\text{Ising}}^{(p)}(\beta)$};
  \node (B) at (5,0) {$Z_{\text{Ising}/\hat{G}}^{(D-p-2)}(\tilde{\beta})$};
  \node (C) at (0,-2) {$Z_{\text{Landau}}^{(p)}(\beta)$};
  \node (D) at (5,-2) {$Z_{\text{Landau}/\hat{G}}^{(D-p-2)}(\tilde{\beta})$};

  \draw[<->] (A) -- node[below] {$\text{KW duality}$} (B);
  \draw[<->] (C) -- node[below] {$\text{KW duality}$} (D);

  \draw[<->] (A) -- node[left] {$\text{HS trans.} $}   (C);
  \draw[<->] (B) --node[right] {$\text{HS trans.}$}  (D);
\end{tikzpicture}
\caption{
The KW duality in lattice higher gauge theories and corresponding Landau field theories.  
The upper KW duality in higher gauge theories is already well established, while the lower one is obtained in this paper by performing the HS transformation.  
In the continuum limit, this leads to an IR duality between continuum   Landau field theories. 
}
\label{duality web}
\end{figure}
Here, IR duality means that two theories describe the same critical phenomena or flow to the same IR critical point by renormalization group.  
It should be noted that we do not intend to provide a rigorous proof of such an IR duality here; rather, we argue that the HS transformation on dual higher gauge theories leads to an exact duality between two lattice Landau theories, which in turn yields an IR duality between two continuum Landau theories at the mean-field level, after taking the classical continuum-limit. 
In Fig.~\ref{duality web}, we summarize these duality relations.~(See the main text for the concrete definitions of partition functions.)

From the perspective of quantum many-body theory of higher dimensional objects, such an IR duality implies that closed extended objects with different dimensions can belong to the same universality class  provided that they share the same higher-form global symmetry.  
%
%
These observations may offer valuable insights into quantum nature of strings and branes.  

\

The organization of this paper is as follows. 
In Section~\ref{sec:2}, we formulate higher gauge theories in a Euclidean $D$-dimensional hypercubic lattice and explain higher-form global symmetries. 
In Section~\ref{sec:Landau}, we derive a Landau field theory for the lattice higher gauge theory by performing the Hubbard-Stratonovich transformations and discuss its continuum-limit.            
We then study the phases of higher gauge theory, including the  explicit  constructions of topological defects in the Landau theory in Section~\ref{sec4}. 
Here, the Coleman-Mermin-Wagner theorem for higher-form global symmetries is also discussed in the present framework.  
In Section~\ref{sec5}, we discuss Kramers-Wannier duality in higher gauge theories and how it can lead to an IR duality in the corresponding continuum Landau theories.   
Section~\ref{sec6} is devoted to summary. 
Appendices provide supplementary details.   

\section{Lattice higher gauge theory}\label{sec:2}

\begin{figure}
 \centering
 \includegraphics[scale=0.3]{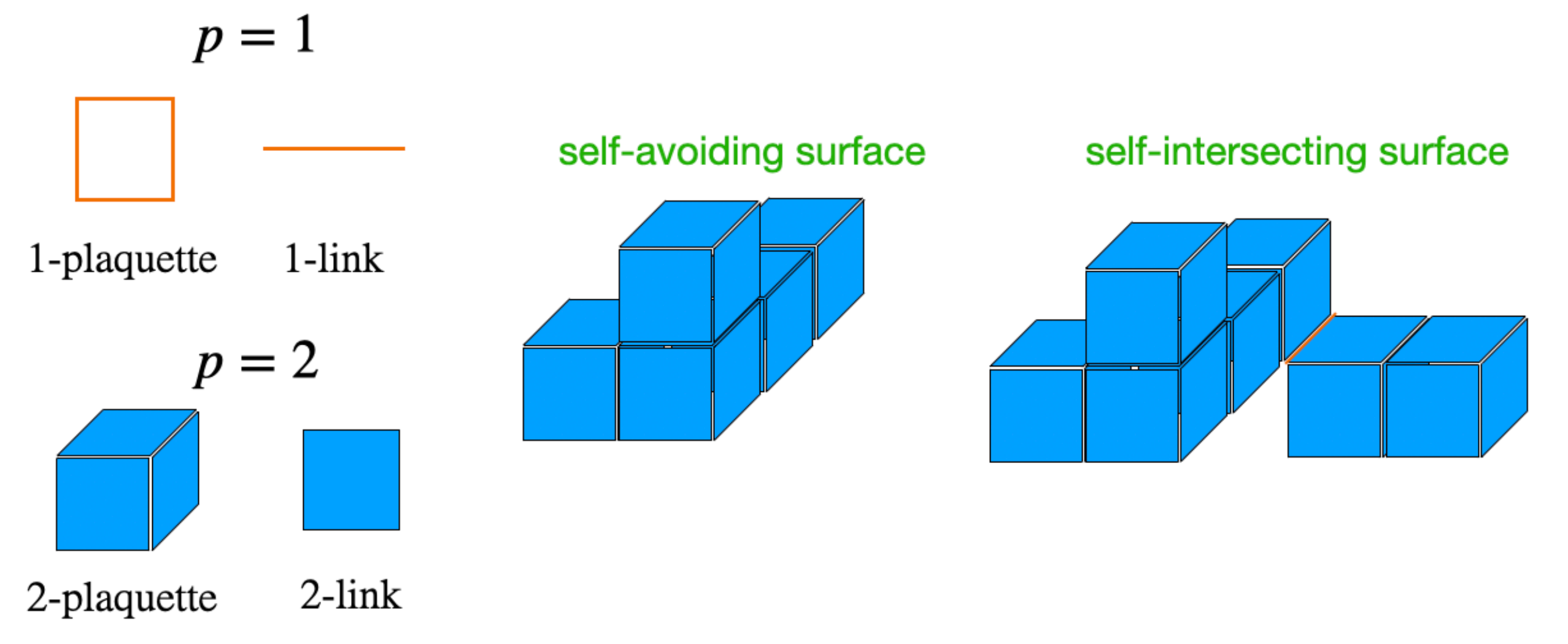}
\caption{
Left: $p$-plaquettes and $p$-links. Here we show $p=1$ and $p=2$ cases. 
    \\
 Right: Examples of closed $2$-dimensional surfaces. 
 The left (right) one is a self-avoiding (intersecting) surface.  
 }
 \label{fig:lattice}
\end{figure}

We consider a $D$-dimensional Euclidean hypercubic lattice $\Lambda_D^{}$ with a lattice spacing $a$.     
%
%
In this discretized space, a $p$-dimensional closed surface $C_p^{}$ is represented as a union of elementary $p$-dimensional minimum hypercubes, as illustrated in the right panel of Fig.~\ref{fig:lattice}.  
Following the same convention as conventional lattice gauge theories, we refer to such a minimum $p$-dimensional hypercube as {\it $p$-link}, and represent it as $L_p^{}$ in general.     
In particular, a $p$-link extending in the $p$-dimensional subspace $(X^{\mu_1^{}},\cdots ,X^{\mu_{p}^{}})$ is explicitly denoted by $L_{\mu_1^{}\cdots \mu_p^{}}^{}(\hat{i})$, where $\hat{i}$ represents the center-of-mass position of it.  

We focus on {\it self-avoiding} $p$-dimensional closed surfaces $C_p^{}$ with orientations, where ``self-avoiding" means that $C_p^{}$ is not a union of multiple $p$-dimensional surfaces that share lower dimensional $q$-links ($q\leq p-1$), as depicted in Fig.~\ref{fig:lattice}. 
In this figure, a $2$-dimensional surface $C_2^{}$ is given by a union of two self-avoiding surfaces sharing a single $1$-link.    
The set of all self-avoiding $p$-dimensional closed surfaces is denoted by $\Gamma_p^{}$.   
Besides, the surface with the opposite orientation to $C_p^{}$ is represented by  $C_p^{-1}\in \Gamma_p^{}$. 

A minimum closed $p$-dimensional surface, called a {\it $p$-plaquette}, is defined by the boundary of a $(p+1)$-link $L_{p+1}^{}$, and is denoted by $\mathrm{P}_p^{}$ in general.  
See the left figures in Fig.~\ref{fig:lattice} for $p=1$ and $2$.       
%
%
%
%
Moreover, we represent the number of $p$-links on a $p$-dimensional surface $C_p^{}\in \Gamma_p^{}$ as $|C_p^{}|$. 
For example, $|\mathrm{P}_p^{}|=2(p+1)$ for ${}^\forall $ $p$-plaquettes $\mathrm{P}_p^{}$.  
In particular, the continuum-limit of a surface $C_p^{}$ is obtained by 
\aln{a\rightarrow 0~,\quad |C_p^{}|\rightarrow \infty~,\quad \text{with}\quad \mathrm{Vol}[C_p^{}]\coloneq a^p |C_p^{}|=\text{fixed}~.
\label{continuum-limit of surface}
} 
Finally, we identify $0$-plaquette as $1$-link, which allows us to discuss lattice field theories with $0$-form global symmetries as well. 
   
\subsection{Higher gauge theory on the lattice}  \label{Higher gauge theory on the lattice}
In the following, we denote a general abelian group as $G$.   
As in ordinary lattice gauge theories, we assign a group variable for each $p$-link as 
%
\aln{
U_{L_{p}^{}}^{}(\hat{i})=\begin{dcases} e^{\frac{2\pi i}{N}m}\quad (m=0,1,\cdots,N-1) & \text{for } G=\mathbb{Z}_N^{}
\\
e^{i\theta}\quad (\theta \in [-\pi,\pi]) & \text{for } G=\mathrm{U}(1)
\end{dcases}.
}
%
As usual, we also define a $p$-link variable corresponding to $L_{p}^{-1}$ as 
\aln{
U_{L_{p}^{-1}}^{}(\hat{i})\coloneqq U^\dagger_{L_{p}^{}}(\hat{i})~.
\label{inverse link}
}
In addition, the gauge transformation of a link variable is defined by  
\aln{
U_{L_{p}^{}}^{}(\hat{i})\quad \rightarrow \quad U_{L_{p}^{}}^{}(\hat{i})\prod_{L_{p-1}^{}\subset \mathrm{P}_{p-1}^{}=\partial L_{p}^{}}g(L_{p-1}^{})~,\quad g(L_{p-1}^{})\in G~, 
\label{gauge transformation}
}
where $\prod_{L_{p-1}^{}\subset \mathrm{P}_{p-1}^{}=\partial L_{p}^{}}$ means the product over all $(p-1)$-links on the boundary $\mathrm{P}_{p-1}^{}=\partial L_{p}^{}$.   
For example, it reproduces the usual gauge transformation 
\aln{
U_{L_{\mu^{}}^{}}^{}(\hat{i})\quad \rightarrow \quad g(i)U_{L_{\mu^{}}^{}}^{}(\hat{i})g^\dagger(i+\hat{\mu})~
} 
when $p=1$.
For a given closed surface $C_p^{}\in \Gamma_p^{}$, we define a Wilson-surface operator by  
\aln{
W[C_p^{}]&\coloneqq \prod_{L_{p}^{}\subset C_p^{}}U_{L_{p}^{}}^{}(\hat{i})~,
\label{path-ordered product}
} 
which is invariant under the gauge transformation~(\ref{gauge transformation}) by definition.   

The partition function of higher gauge theory defined on $p$-links is   
\aln{
Z_{\text{Ising}}^{(p)}(\beta)&=\int [dU]\exp\left(-S_\mathrm{E}^{}\right)~,\label{lattice partition function}
\\
S_\mathrm{E}^{}&=\beta \sum_{\mathrm{P}_p^{}}\left[1-\frac{1}{2}\left(W[\mathrm{P}_p^{}]+W[\mathrm{P}_p^{-1}]\right)\right]~,
\label{lattice Ising model}
}
where $\beta=1/g^2$ is the inverse gauge coupling, $\sum_{\mathrm{P}_p^{}}$ is the summation over all positive oriented $p$-plaquettes, and 
\aln{
\int [dU]\coloneq  
\begin{dcases}\prod_{L_{p}^{}}\left(\frac{1}{|G|}\sum_{g(L_{p}^{})\in G}\right)~
& \text{for a finite abelian group $G$}
\\
\prod_{L_{p}^{}}\frac{1}{2\pi}\int_{-\pi}^{\pi} d\theta_{L_{p}^{}}^{}
& \text{for $G=\mathrm{U}(1)$}
\end{dcases}
\label{measure}
}
is the path-integral measure of link variables. 
For $p=0$, Eqs.~(\ref{lattice partition function})(\ref{lattice Ising model}) reduce to an Ising model because $0$-plaquette is identified as $1$-link.   
In this sense, the lattice gauge theory~(\ref{lattice partition function}) can be viewed as a higher-link version of Ising models.  
This is the reason why we put the subscript ``Ising" in the partition function~(\ref{lattice partition function}).
%

\subsection{Higher form global symmetry}

\begin{figure}
    \centering
     \includegraphics[scale=0.3]{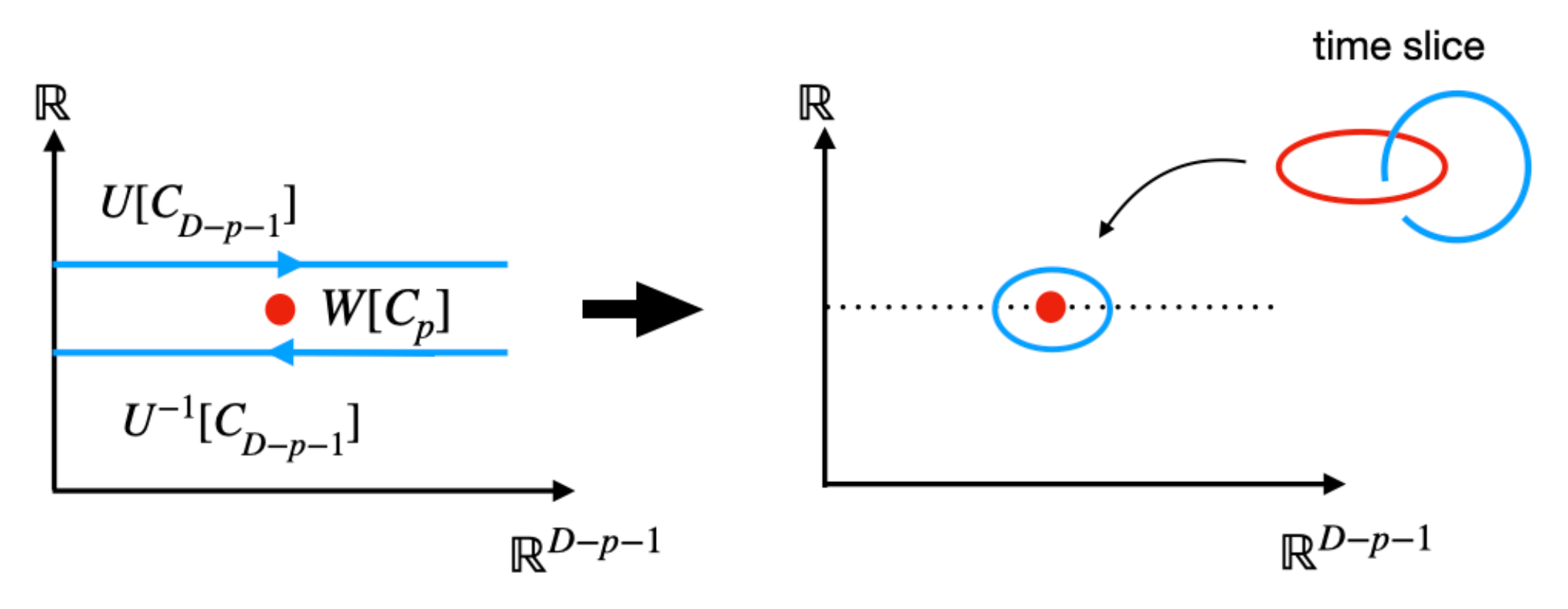}
    \caption{
  Graphical representation of the operator relation~(\ref{operator relation}). 
  Here, the red point corresponds to the Wilson-surface operator $W[C_p^{}]$, and the blue lines correspond to the symmetry operators $U[C_{D-p-1}^{}],~U[C_{D-p-1}^{}]^{-1}$.   
    }
    \label{fig:operator}
\end{figure}
%
%
The lattice gauge theory~(\ref{lattice partition function}) is invariant under the following $p$-form global transformation 
\aln{
W[C_p^{}]\quad \rightarrow \quad g(C_p^{})W[C_p^{}]~,\quad g(C_p^{})\in G~,
\label{p-form symmetry}
}
with the condition $g(C_p^{})=1$ for all $C_p^{}\in \Gamma_p^{}$ that shrink to a point in the continuum-limit~(\ref{continuum-limit of surface}). 
This condition  ensures the local flatness of gauge parameters in the continuum-limit.  
Besides, $g(C_p^{})$ satisfies 
\aln{g(C_p^{1}+C_p^{2})=g(C_p^{1})g(C_p^{2})~
\label{composition rule}
}
for a combined surface $C_p^{1}+C_p^2\in \Gamma_p^{}$ because $W[C_p^1+C_p^2]=W[C_p^1]W[C_p^2]$. 
%
More explicitly, the $p$-form transformation is expressed by 
\aln{
g(C_p^{})=\begin{dcases}
e^{i\theta\int_{C_p^{}}\Lambda_p^{}}~,\quad \theta \in \mathbb{R} & \text{for }G=\mathrm{U}(1)
\\e^{\frac{i}{N}\int_{C_p^{}}\Lambda_p^{}}. & \text{for }G=\mathbb{Z}_N^{}
\end{dcases},\quad 
 d\Lambda_p^{}=0~,\quad \int_{C_p^{}}\Lambda_p^{}\in 2\pi \mathbb{Z}~,
} 
in the continuum-limit expression, where $\Lambda_p^{}$ is a differential $p$-form. 
%
%
The corresponding symmetry/topological operator is a codimension $(p+1)$ operator $U[C_{D-p-1}^{}]$ satisfying 
\aln{
U [C_{D-p-1}^{}]W[C_p^{}]U^{-1}[C_{D-p-1}^{}]=g(C_p^{})^{\mathrm{I}[C_p^{},C_{D-p-1}^{}]}W[C_p^{}]~,
\label{operator relation}
}
as an (equal-time) operator relation, where $C_{D-p-1}^{}$ is a $(D-p-1)$-dimensional subspace and $\mathrm{I}[C_p^{},C_{D-p-1}^{}]$ is the intersection number between $C_p^{}$ and $C_{D-p-1}^{}$. 
See Fig~\ref{fig:operator} for a graphical representation of this relation.

For $G=\mathrm{U}(1)$, the symmetry operator is explicitly given by 
\aln{
U_\theta^{}[C_{D-p-1}^{}]=\exp\left(i\theta\int_{C_{D-p-1}^{}}\star F_{p+1}^{}\right)~,\quad F_{p+1}^{}=dA_p^{}
}  
in the continuum-limit expression.
When $G$ is a finite abelian group, it can be realized as a locally flat configuration of the background $(p+1)$-form gauge field coupled to Eq.~(\ref{lattice partition function})~\cite{Gaiotto:2014kfa}. 
%
See also Appendix~\ref{app:operator relation} for the explicit check of Eq.~(\ref{operator relation}) in the continuum higher gauge theories.     

\section{Landau theory for higher gauge theory}\label{sec:Landau}

We derive a Landau field theory for the lattice higher gauge theory~(\ref{lattice partition function}) and examine its (classical) continuum-limit. 
The key step in the derivation is to rewrite the lattice action~(\ref{lattice Ising model}) into a quadratic form by introducing an infinitesimal deformation operator of closed surfaces. 
Such a quadratic form enables us to perform the Hubbard-Stratonovich (HS) transformation, and we obtain a scalar field theory defined on the space of self-avoiding closed surfaces $\Gamma_p^{}$.  

\subsection{Hubbard Stratonovich transformation}\label{sec:dual transformation}

\begin{figure}
    \centering
     \includegraphics[scale=0.25]{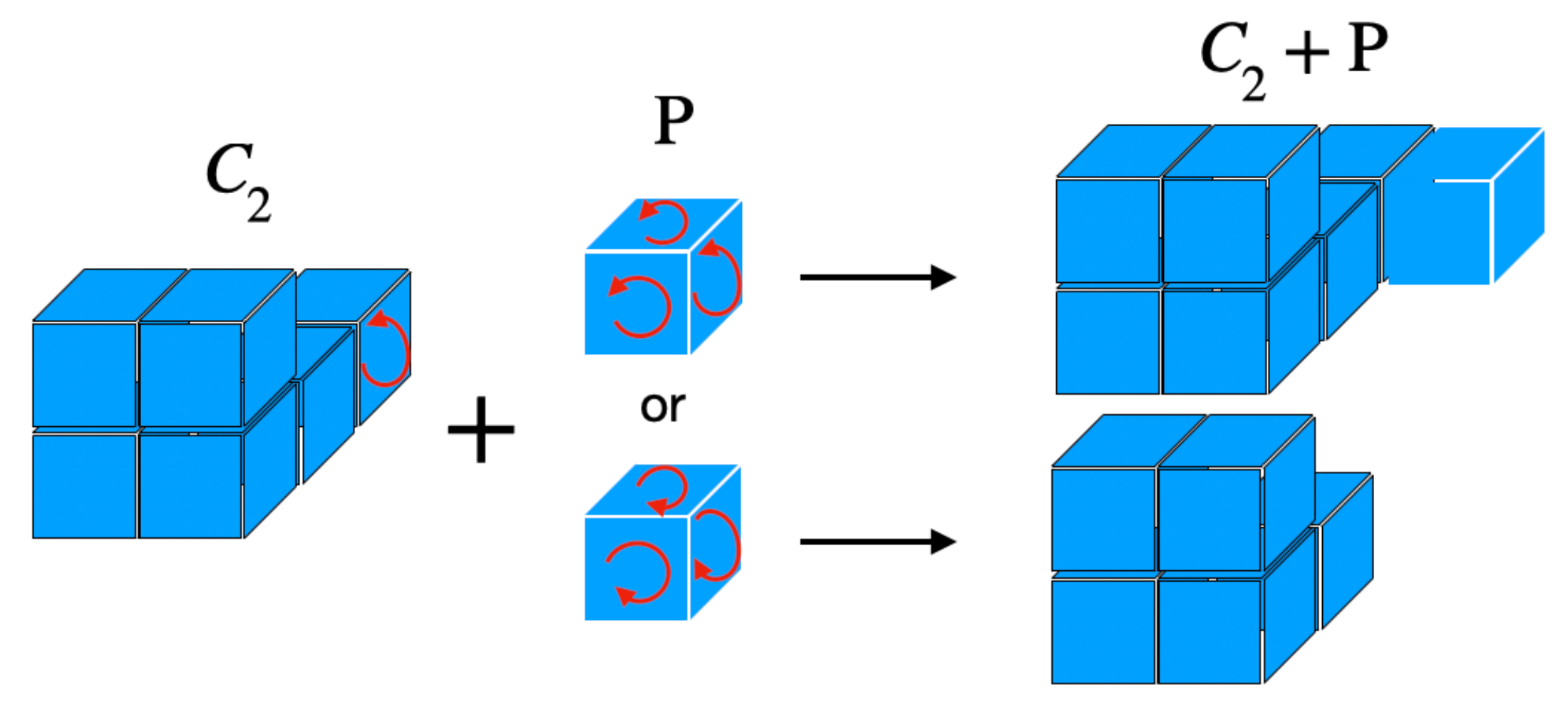}
    \caption{Combination of a $2$-dimensional surface $C_2^{}$ and $2$-plaquette $\mathrm{P}_2^{}$. 
    In the upper case, a single $2$-link (face) is erased and the remaining five  $2$-links are attached to $C_2^{}$.
    In the lower case, on the other hand, five $2$-links are erased and the remaining one $2$-link is attached.  
    }
    \label{fig:sum}
\end{figure}

%
Given a self-avoiding surface $C_p^{}\in \Gamma_p^{}$ and a $p$-plaquette $\mathrm{P}_p^{}$, we define a combined surface $C_p^{}+\mathrm{P}_p^{}$ by erasing the common $p$-links (with opposite orientations) between $C_p^{}$ and $\mathrm{P}_p^{}$, and attaching the remaining $p$-links of $\mathrm{P}_p^{}$ which are not common with $C_p^{}$, as illustrated in Fig.~\ref{fig:sum}. 
%
%
%
An trivial but important property is 
\aln{
W[C_p^{}]^* W[C_p^{}+\mathrm{P}_p^{}]=W[\mathrm{P}_p^{}] 
\label{U property}
}
when $C_p^{}+\mathrm{P}_p^{}\in \Gamma_p^{}$. 
When $C_p^{}+\mathrm{P}_p^{}\notin \Gamma_p^{}$, Eq.~(\ref{U property}) does not necessarily hold.

We introduce a translation operator along a $p$-laquette $\mathrm{P}_p^{}$  by~\cite{Yoneya:1980bw,Banks:1980sq} 
\aln{
\Pi_{\mathrm{P}_p^{}}^{}W[C_p^{}]\coloneq \begin{cases}W[C_p^{}+\mathrm{P}_p^{}] & \text{for } C_p^{}+\mathrm{P}_p^{}\in \Gamma_p^{}
\\
0 & \text{for } C_p^{}+\mathrm{P}_p^{}\notin \Gamma_p^{}
\end{cases}~.
\label{translation operator}
}
Using this operator, we can express the lattice action~(\ref{lattice Ising model}) as
\aln{
S_\mathrm{E}^{}&=\text{constant}-\frac{\beta}{2}\sum_{C_{p}^{}\in \Gamma_p^{}}w[C_p^{}]W[C_p^{}]^* (k_0^{}+\hat{H})w[C_p^{}]^* W[C_p^{}]
~,
\label{quadratic action}
}
where the operator $\hat{H}$ is defined by 
\aln{
\hat{H}W[C_p^{}]\coloneqq \frac{1}{2(D-p)|C_p^{}|}\sum_{\substack{\mathrm{P}_p^{}\\\mathrm{P}_p^{}\cap C_p^{}\neq \emptyset}}\Pi_{\mathrm{P}_p^{}}^{}W[C_p^{}]~.
\label{derivative operator}
}
We will soon prove the equivalence below. 
Here, $\sum_{\substack{\mathrm{P}_p^{}\\\mathrm{P}_p^{}\cap C_p^{}\neq \emptyset}}$ implies the summation over all $p$-plaquettes having common $p$-links with $C_p^{}$, and $k_0^{}$ is a sufficiently large positive constant to make $k_0^{}+\hat{H}$ positive definite.  
The introduction of $k_0^{}$ does not change any observables because it simply  corresponds to a constant shift of the action.   
Note also that the normalization of $\hat{H}$ is chosen to guarantee $\hat{H}\rightarrow 1$ in the continuum-limit $a\rightarrow 0$.   
See the next subsection for more details. 

Moreover, $w[C_p^{}]$ is an arbitrary weight functional that is invariant under spacetime translation and satisfies 
\aln{
\frac{1}{2(D-p)}\sum_{\substack{C_p^{}\in \Gamma_p^{}\\C_p^{}\cap \mathrm{P}_p^{}\neq \emptyset}}\frac{1}{|C_p^{}|}w[C_p^{}]w[C_p^{}+\mathrm{P}_p^{}]^*=1~,
\label{w normalization}
}
%
where $\sum_{\substack{C_p^{}\in \Gamma_p^{}\\C_p^{}\cap \mathrm{P}_p^{}\neq \emptyset}}$ means the summation over all closed surfaces $C_p^{}\in \Gamma_p^{}$ such that (i) they have common $p$-links with a given $p$-plaquette $\mathrm{P}_p^{}$ and (ii) $C_p^{}+\mathrm{P}_p^{}$ also belongs to $\Gamma_p^{}$.    
A simplest choice satisfying these conditions is
\aln{
w[C_p^{}]= b\times \exp\left(-\frac{\alpha}{2}\mathrm{Vol}[C_p^{}]+i\int_{C_p^{}}\lambda_p^{}\right)~,
\label{simple weight function}
}
where $\lambda_p^{}$ is a differential closed $p$-form $d\lambda_p^{}=0$ with $\int_{C_p^{}} \lambda_p^{}\in 2\pi \mathbb{Z}$, and $\alpha>0$ is a brane tension with mass dimension $p$.   
The normalization factor $b$ is determined by Eq.~(\ref{w normalization}) as 
\aln{
\frac{b^2}{2(D-p)}\sum_{\substack{C_p^{}\in \Gamma_p^{}\\C_p^{}\cap \mathrm{P}_p^{}\neq \emptyset}}\frac{1}{|C_p^{}|}&e^{-\frac{\alpha}{2}(\mathrm{Vol}[C_p^{}]+\mathrm{Vol}[C_p^{}+\mathrm{P}_p^{}])-i\int_{\mathrm{P}_p^{}}^{}\lambda_p^{}}\approx \frac{b^2}{2(D-p)}\sum_{\substack{C_p^{}\in \Gamma_p^{}\\C_p^{}\cap \mathrm{P}_p^{}\neq \emptyset}}\frac{1}{|C_p^{}|}e^{-\alpha \mathrm{Vol}[C_p^{}]}=1
\nn
&\therefore~b^2\approx \frac{2(D-p)}{\sum_{\substack{C_p^{}\in \Gamma_p^{}\\C_p^{}\cap \mathrm{P}_p^{}\neq \emptyset}}e^{-\alpha \mathrm{Vol}[C_p^{}]}/|C_p^{}|}~,
}
where we have used the Stokes theorem in the first line.  
The volume-suppression factor in Eq.~(\ref{simple weight function}) was introduced  in order to make the summation over $p$-dimensional surfaces well-defined.   
Now we can prove the equivalence between Eq.~(\ref{lattice Ising model}) and (\ref{quadratic action}) as follows: 
\aln{
\sum_{C_p^{}\in \Gamma_p^{}} w[C_p^{}]W^*[C_p^{}]\hat{H}w[C_p^{}]^* W[C_p^{}]
&=\sum_{C_p^{}\in \Gamma_p^{}}w[C_p^{}]W^*[C_p^{}]\frac{1}{2(D-p)|C_p^{}|}\sum_{\substack{\mathrm{P}_p^{}\\ \mathrm{P}_p^{}\cap C_p^{}\neq \emptyset}}w[C_p^{}+\mathrm{P}_p^{}]^* W[C_p^{}+\mathrm{P}_p^{}]
\nn
&=\sum_{\mathrm{P}_p^{},\mathrm{P}_p^{-1}}W[\mathrm{P}_p^{}]\frac{1}{2(D-p)}\sum_{\substack{C_p^{}\in \Gamma_p^{}\\ C_p^{}\cap \mathrm{P}_p^{}\neq \emptyset}}\frac{1}{|C_p^{}|}w[C_p^{}]w[C_p^{}+\mathrm{P}_p^{}]^*
\nn
&=\sum_{\mathrm{P}_p^{}}(W[\mathrm{P}_p^{}]+W[\mathrm{P}_p^{-1}])~,
\label{proof}
}
where we have used Eq.~(\ref{U property}) (Eq.~(\ref{w normalization})) from the first (second) line to the second (third) line. 
Note also that $\hat{H}$ is self-adjoint as 
\aln{
\sum_{C_p^{}\in \Gamma_p^{}}\phi_1^*[C_p^{}]\hat{H}\phi_2^{}[C_p^{}]&=\sum_{C_p^{}\in \Gamma_p^{}}\frac{1}{2(D-p)|C_p^{}|}
\sum_{\substack{\mathrm{P}_p^{}\\ \mathrm{P}_p^{}\cap C_p^{}\neq \emptyset}}\phi_1^*[C_p^{}]\phi_2^{}[C_p^{}+\mathrm{P}_p^{}]
\nn
&=\sum_{\mathrm{P}_p^{}}\sum_{\substack{C_p^{}\in \Gamma_p^{}\\C_p^{}\cap \mathrm{P}_p^{}\neq \emptyset}} \frac{1}{2(D-p)|C_p^{}|}\phi_1^*[C_p^{}]\phi_2^{}[C_p^{}+\mathrm{P}_p^{}]~.
\label{self adjointness}
}
%
Since $C^{'}_p \coloneqq C_p^{}+\mathrm{P}$ belongs to $\Gamma_p^{}$ by the definition of $\Pi_P^{}$, Eq.~(\ref{self adjointness}) can be also written as
\aln{
=\sum_{\mathrm{P}_p^{}} \sum_{\substack{C_{p}^{'}\in\Gamma_p^{}\\C_p^{'}\cap \mathrm{P}_p^{}\neq \emptyset}}\frac{1}{2(D-p)|C_p^{}|}\phi_1^*[C'_p+\mathrm{P}_p^{-1}]\phi_2^{}[C'_p]=\sum_{C_p^{}\in \Gamma_p^{}}\left\{\hat{H}\phi_1^\dagger[C_p^{}]\right\}\phi_2^{}[C_p^{}]~,
}
which confirms $\hat{H}^\dagger=\hat{H}$. 

Since Eq.~(\ref{quadratic action}) is a quadratic form, we can perform the HS transformation and obtain  
\aln{
Z^{(p)}_{\text{Ising}}(\beta)&\propto \int [dU]\int [d\phi]\exp\bigg\{-\sum_{C_p^{}\in \Gamma_p^{}}\left(\frac{2}{\beta}\phi[C_p^{}]^*(k_0^{}+\hat{H})^{-1}\phi[C_p^{}]+w[C_p^{}]^*W[C_p^{}]
\phi[C_p^{}]+{\rm h.c.}
\right)
\bigg\}
\\
&=\int [d\phi]\exp\left\{-\sum_{C_p^{}\in \Gamma_p^{}}\left(\frac{2}{\beta}\phi[C_p^{}]^* (k_0^{}+\hat{H})^{-1}\phi[C_p^{}]+V(\phi,\phi^*)\right)\right\}~,
\label{loop functional theory}
} 
where $\phi[C_p^{}]\in \mathbb{C}$ is a functional field on the space of self-avoiding $p$-dimensional closed surfaces $\Gamma_p^{}$, and  
\aln{
\sum_{C_p^{}\in \Gamma_p^{}}V(\phi,\phi^*)=-\log\left[\int [dU]\exp\left(-\sum_{C_p^{}\in \Gamma_p^{}}w[C_p^{}]W[C_p^{}]^*\phi[C_p^{}]+{\rm h.c.}\right)\right]~
\label{dual potential}
}
is the potential of $\phi[C_p^{}]$. 
One can see that the lattice higher gauge theory is now recast as a scalar field theory defined on $\Gamma_p^{}$.  

\

In this functional formulation, the $p$-form global transformation~(\ref{p-form symmetry}) is expressed by 
\aln{
\phi[C_p^{}]\quad \rightarrow \quad g(C_p^{})\phi[C_p^{}]~,\quad g(C_p^{})\in G~.
\label{scalar p-form transformation}
}
Although the invariance of the potential~(\ref{dual potential}) under this transformation is trivial, the invariance of the kinetic term may be  nontrivial. 
By the definition of $\hat{H}$~(\ref{derivative operator}), we can see
\aln{
g(C_p^{})^*\hat{H}g(C_p^{})&=\frac{1}{2(D-p)|C_p^{}|}\sum_{\substack{\mathrm{P}_p^{}\\\mathrm{P}\cap C_p^{}\neq \emptyset}}g(C_p^{})^*g(C_p^{}+\mathrm{P}_p^{})
\nn
&=\frac{1}{2(D-p)|C_p^{}|}\sum_{\substack{\mathrm{P}_p^{}\\\mathrm{P}_p^{}\cap C_p^{}\neq \emptyset}}g(\mathrm{P}_p^{})=1~,
} 
where we have used Eq.~(\ref{composition rule}) and $g(\mathrm{P}_p^{})=1$.\footnote{
For a given $p$-link $L_p^{}\subset C_p^{}$, there are $2(D-p)$ $p$-plaquettes that contain $L_p^{}$, which leads to $\sum_{\substack{\mathrm{P}_p^{}\\ \mathrm{P}_p^{}\cap C_p^{}\neq \emptyset}}1=2(D-p)|C_p^{}|$. 
}
This proves the invariance of the kinetic term under Eq.~(\ref{scalar p-form transformation}). 

In general, the invariance under the $p$-form transformation~(\ref{scalar p-form transformation}) restricts the functional form of $V(\phi,\phi^*)$. 
For example, one can generally expand it as 
\aln{
\sum_{C_p^{}\in \Gamma_p^{}}V(\phi,\phi^*)=&\sum_{C_p^{}\in \Gamma_p^{}}\left\{\mu_2^{}(C_p^{})\phi[C_p^{}]^*\phi[C_p^{}]+\mu_4^{}(C_p^{})(\phi[C_p^{}]^*\phi[C_p^{}])^2+\cdots 
\right\}
\nn
+\sum_{C_p^{1}\in \Gamma_p^{}}&\sum_{C_p^{2}\in \Gamma_p^{}}\sum_{C_p^{3}\in \Gamma_p^{}}\delta(C_p^{1}-C_p^{2}-C_p^{3})\mu_3^{}(C_p^1,C_p^2,C_p^3) \phi[C_p^{1}]^* \phi[C_p^{2}]\phi[C_p^{3}]+\text{h.c.}+\cdots~,
\label{potential expansion}
}
where $\delta (C_p^{})$ is the delta function on $\Gamma_p^{}$, i.e.,
\aln{
\sum_{{C'}_p^{}\in \Gamma_p^{}}\delta(C_p^{}-{C'}_p^{})\phi[C_p^{'}]=\phi[C_p^{}]~,
}
for an arbitrary functional $\phi[C_p^{}]$.  
The existence of this delta function in the second line in Eq.~(\ref{potential expansion}) ensures the invariance under the $p$-form global transformation as 
\aln{
 \delta(C_p^{1}-C_p^{2}-C_p^{3}) g(C_p^{1})^*g(C_p^{2})g(C_p^{3})&= \delta(C_p^{1}-C_p^{2}-C_p^{3})g(C_p^{1}-C_p^{2}-C_p^3)
\nn
&=\delta(C_p^{1}-C_p^{2}-C_p^{3})~,
\label{cancellation of g}
}
where we have used Eq.~(\ref{composition rule}).  
The same argument applies to higher-order non-local interactions in general.    
We should note that the coupling constants $\{\mu_i^{}(C_p^{})\}$ in the expansion~(\ref{potential expansion}) can depend on $C_p^{}$ for a general weight functional $w[C_p^{}]$.   
For the simplest example~(\ref{simple weight function}), however, they  become constant as
\aln{
\begin{cases}\mu_2^{}(C_p^{})&\propto w[C_p^{}]^* w[C_p^{}]=b^2
\\
\mu_3^{}(C_p^1,C_p^2,C_p^3)&\propto w[C_p^{1}]^* w[C_p^{2}]w[C_p^{3}]=b^3
\end{cases}
}
by absorbing the volume suppression factor $e^{-\alpha \mathrm{Vol}[C_p^{}]}$ into  the Wilson-surface operator as $e^{-\alpha \mathrm{Vol}[C_p^{}]}W[C_p^{}]\rightarrow W[C_p^{}]$ in Eq.~(\ref{dual potential}).  
%
%
Furthermore, when $G$ is a finite abelian group, the potential can also contain  explicit $\mathrm{U}(1)$ breaking terms.  
For example, for $G=\mathbb{Z}_N^{}$, there can be   
\aln{
\sum_{C_p^{}\in \Gamma_p^{}}V(\phi,\phi^*)\ni \sum_{C_p^{}\in \Gamma_p^{}}\left\{\alpha^{(1)}(C_p^{})\phi[C_p^{}]^N+\alpha^{(2)}(C_p^{})(\phi[C_p^{}]^N)^2+\cdots \right\}+{\rm h.c.}~,
\label{determinant terms}
}
where the couplings $\{\alpha^{(i)}(C_p^{})\}$ transform trivially under the $\mathbb{Z}_N^{}$ transformation~(\ref{scalar p-form transformation}).   
For a general finite abelian group $G$, the structure of the potential becomes more complicated.

%
%
\

One more comment is in order. 
In Eq.~(\ref{loop functional theory}), the canonically normalized kinetic term can be realized by the field redefinition $g\phi[C_p^{}]\rightarrow \phi[C_p^{}]$, which in turn leads to $g^{-n}\phi[C_p^{}]^n$ in a general $n$-th order term in the potential $V(\phi,\phi^*)$. 
This implies that the weak-coupling limit in the gauge theory $g^{}\rightarrow 0$ corresponds to the strong coupling regime $g^{-n}\rightarrow \infty$ in the Landau field theory, and vice versa. 
In Section~\ref{sec4}, we will see that this correspondence plays an important role to demonstrate the area/perimeter law in the mean-field analysis within the Landau theory. 
%

\subsection{Classical continuum limit}\label{sec3}
Let us discuss classical continuum-limit of the Landau theory~(\ref{loop functional theory}).  
We represent a $D$-dimensional Euclidean curved space with a metric $g_{\mu\nu}^{}(X)$ by $\Sigma_D^{}$.  
In the continuum limit $a\rightarrow 0$, a self-avoiding surface $C_p^{}\in \Gamma_p^{}$ is represented by a set of embedding functions 
\aln{
X^\mu: S^p\quad \rightarrow \quad \Sigma_D^{}~,\quad (\mu=1,2,\cdots,D)~,
}
where $S^p$ is the $p$-dimensional sphere whose general intrinsic coordinates are denoted by $\{\xi^i\}_{i=1}^p$.   
Then, the induced metric on $C_p^{}$ is  
\aln{
ds^2=g_{\mu\nu}^{}(X)dX^\mu dX^\nu=g_{\mu\nu}^{}(X(\xi))\frac{\partial X^\mu}{\partial\xi^i} \frac{\partial X^\nu}{\partial\xi^j}d\xi^id\xi^j= h_{ij}^{}(\xi)d\xi^id\xi^j~,
}
and we denote its determinant as $h(\xi)\coloneq \mathrm{det}(h_{ij}^{}(\xi))\geq 0$. 
Then, the volume of $C_p^{}$ is expressed as 
\aln{
\mathrm{Vol}[C_p^{}]=a^p|C_p^{}|\quad \rightarrow \quad \int_{S^p}^{}d^p\xi \sqrt{h(\xi)}
}
in the continuum-limit~(\ref{continuum-limit of surface}). 
This is invariant under the reparametrization $\xi^i~\rightarrow ~{\xi'}^i=f^i(\xi)$ on $C_p^{}$.  
This can be also expressed in a differential form as 
\aln{\mathrm{Vol}[C_p^{}]=\int_{C_p^{}}E_p^{}~, 
}
with
\aln{E_{p}^{}&=\frac{1}{p!}(E_{p}^{}(\xi))^{\mu_{1}^{}\cdots \mu_{p}^{}}dX_{\mu_{1}^{}}^{}\wedge\cdots \wedge dX_{\mu_{p}^{}}^{}
\coloneqq\frac{1}{p!\sqrt{h(\xi)}}\{X^{\mu_1^{}},\cdots,X^{\mu_p^{}}\}dX_{\mu_1^{}}^{}\wedge \cdots \wedge dX_{\mu_{p}^{}}^{}~,
}
where 
\aln{\{X^{\mu_1^{}},\cdots,X^{\mu_p^{}}\}=\epsilon^{i_1^{}\cdots i_p^{}}\frac{\partial X^{\mu_1^{}}}{\partial \xi^{i_1^{}}}\cdots \frac{\partial X^{\mu_p^{}}}{\partial \xi^{i_p^{}}}
}
is the Nambu-bracket~\cite{Nambu:1973qe}. 
Geometrically, $E_p^{}$ corresponds to the normalized infinitesimal volume element
 on $C_p^{}$, which implies that the discretized version of the component $(E_{p}^{}(\xi))^{\mu_1^{}\cdots \mu_p^{}}$ is 
\aln{
(E_{p}^{}(\hat{i}))^{\mu_1^{}\cdots \mu_p^{}}=\frac{1}{p!}\sum_{\sigma\in S_p^{}}\delta^{\sigma(\mu_1^{})}_{\nu_1^{}}\cdots \delta^{\sigma(\mu_p^{})}_{\nu_p^{}}~
\label{discretized Nambu-bracket}
} 
for $L_{\nu_1^{}\cdots \nu_p^{}}(\hat{i})\subset C_p^{}$. 
Here, $S_p^{}$ is the symmetric group of degree $p$. 
%
%
%
For the later convenience, we also represent the volume of the $(p+1)$-dimensional  minimal surface $M_{p+1}^{}$ bounded by $C_p^{}$ as 
\aln{\mathrm{Vol}[M_{p+1}^{}]\coloneq \int_{M_{p+1}}^{}E_{p+1}^{}~.
}

\

With these preparations, we can study the classical continuum-limit of Eq.~(\ref{loop functional theory}). 
The most crucial and nontrivial point is the treatment of the kinetic term $(k_0^{}+\hat{H})^{-1}$, which is constructed by the plaquette shifting operator~(\ref{translation operator}). 
This fact motivates us to generalize the notion of ``derivative" from the ordinary one $\partial/\partial x^{\mu}$ to a functional one acting on functional fields $\phi[C_p^{}]$.     
In fact, such a generalization has been investigated for the construction of   effective Landau field theories with higher-form global symmetries~\cite{Iqbal:2021rkn,Hidaka:2023gwh,Kawana:2024fsn,Kawana:2024qmz,Kawana:2025vvf}.    
In the present lattice formulation, it can be explicitly defined by
\aln{
\frac{\delta \psi[C_p^{}]}{\delta \sigma^{\mu_1^{}\cdots \mu_{p+1}}(\hat{i})}\coloneqq \frac{\psi[C_p^{}+\mathrm{P}_{\mu_1^{}\cdots \mu_{p+1}}^{}(\hat{i})]-\psi[C_p^{}]}{a^{p+1}} 
\label{def of area derivative}
}
for a general functional field $\psi[C_p^{}]$ and a general $p$-dimensional surface $C_p^{}$. 
(That is, it does not need to be a self-avoiding surface.)
Here, $\mathrm{P}_{\mu_1^{}\cdots \mu_{p+1}}^{}(\hat{i})$ is a $p$-plaquette sitting in the $(p+1)$-dimensional subspace $(X^{\mu_1^{}},\cdots,X^{\mu_{p+1}^{}})$ with a center-of-mass position $\hat{i}$.  
Since a permutation of the target-space coordinates corresponds to a change of orientation of $C_p^{}$, we can naturally define
\aln{
\mathrm{P}_{\sigma(\mu_1^{})\cdots \sigma(\mu_{p+1}^{})}^{}(\hat{i})=\begin{cases} \mathrm{P}_{\mu_1^{}\cdots \mu_{p+1}^{}}^{}(\hat{i}) & \text{for $\mathrm{sgn}(\sigma)=1$}
\\
 \mathrm{P}_{\mu_1^{}\cdots \mu_{p+1}^{}}^{-1}(\hat{i})
 & \text{for $\mathrm{sgn}(\sigma)=-1$}
\end{cases},
} 
where $\sigma$ is an element of the symmetric group $S_{p+1}^{}$ and $\mathrm{sgn}(\sigma)$ is the sign of the permutation. 
This leads to the anti-symmetric property of the area derivatives~\cite{Iqbal:2021rkn,Hidaka:2023gwh,Kawana:2024fsn,Kawana:2024qmz,Kawana:2025vvf}:
\aln{
\frac{\delta}{\delta \sigma^{\sigma(\mu_1^{})\cdots \sigma(\mu_{p+1}^{})}(\hat{i})}=\mathrm{sgn}(\sigma)\frac{\delta}{\delta \sigma^{\mu_1^{}\cdots \mu_{p+1}^{}}(\hat{i})}~.
}   
Note that the center-of-mass position $\hat{i}$ corresponds to a point $\xi=(\xi^1,\cdots,\xi^{p})$ on $C_p^{}$ in the continuum-limit $a\rightarrow 0$. 
%

One of the important properties of the area derivatives is~\cite{Iqbal:2021rkn,Hidaka:2023gwh}
\aln{
\frac{\delta}{\delta \sigma^{\mu_1^{}\cdots \mu_{p+1}^{}}(\xi)}\left(\int_{C_p^{}} B_p^{}\right)= (dB_p^{})_{\mu_1^{}\cdots \mu_{p+1}^{}}^{}(X(\xi))
\label{property of area derivative}
}
for an arbitrary differential $p$-form $B_p^{}(X)$, which then allows us to  calculate the area derivative on an arbitrary function of $\int_{C_p^{}} B_p^{}$ as 
\aln{
\frac{\delta}{\delta \sigma^{\mu_1^{}\cdots \mu_{p+1}^{}}(\xi)}f\left(\int_{C_p^{}} B_p^{}\right)=(dB_p^{})_{\mu_1^{}\cdots \mu_{p+1}^{}}^{}(X(\xi))f'(x)\bigg|_{x=\int_{C_p^{}} B_p^{}}~. 
}
This can be interpreted as a generalization of the chain rule of ordinary derivative $\partial_\mu^{}$.

In the Landau field theory~(\ref{loop functional theory}), we must take into account the fact that the translation operator $\Pi_{\mathrm{P}_p^{}}^{}$ is nonzero only for $p$-plaquettes that share common $p$-links with $C_p^{}$.     
This corresponds to projecting the area derivative~(\ref{def of area derivative}) onto the direction of a shared $p$-link as 
\aln{
\frac{\delta}{\delta S^\mu(\hat{i})}\coloneq (E_{p}^{}(\hat{i}))^{\nu_1^{}\cdots \nu_p^{}}\frac{\delta}{\delta \sigma^{\mu\nu_1^{}\cdots \nu_p^{}}(\hat{i})}~,
\label{projection}
}
where the projection tensor $(E_{p}^{}(\hat{i}))^{\nu_1^{}\cdots \nu_p^{}}$ is given by Eq.~(\ref{discretized Nambu-bracket}). 
%
%

Now we can evaluate $\hat{H}$ for $a\rightarrow 0$ as 
\aln{
\hat{H}\phi[C_p^{}]&=\frac{1}{2(D-p)|C_p^{}|}\sum_{\substack{\mathrm{P}_p^{}\\ \mathrm{P}_p^{}\cap C_p^{}\neq \emptyset}}\Pi_{\mathrm{P}_p^{}}^{}\phi[C_p^{}]=\phi[C_p^{}]+\frac{a^{p+1}}{2(D-p)|C_p^{}|}\sum_{\hat{i}\in C_p^{}}\sum_{\mu=1}^D\sum_{s=\pm 1} 
\mathrm{sgn}(s)\frac{\delta \phi[C_p^{}]}{\delta S^{\mu}(\hat{i})}
\nn
&\hspace{4cm}+\frac{2(a^{p+1})^2}{4(D-p)|C_p^{}|}\sum_{\hat{i}\in C_p^{}}\sum_{\mu=1}^D
\frac{\delta \phi[C_p^{}]}{\delta S^{\mu}(\hat{i})\delta S^{\mu}(\hat{i})}+{\cal O}(a^{3(p+1)})
\\
&=\phi[C_p^{}]+\frac{(a^{p+1})^2}{2(D-p)|C_p^{}|}\sum_{\hat{i}\in C_p^{}}
\sum_{\mu=1}^D\frac{\delta^2 \phi[C_p^{}]}{\delta S^{\mu}(\hat{i})\delta S^{\mu}(\hat{i})}+{\cal O}(a^{3(p+1)})~,
}
where the linear terms in $a^{p+1}$ vanish due to the contributions with opposite orientations.   
%
Then, using the Einstein notation, we obtain 
\aln{
\lim_{a\rightarrow 0}\phi[C_p^{}]^*\frac{\hat{H}^{}-1}{(a^{p+1})^2}\phi[C_p^{}]
&=\frac{1}{2(D-p)|C_p^{}|}\lim_{a\rightarrow 0}\sum_{\hat{i}\in C_p^{}}
\phi[C_p^{}]^*\frac{\delta^2}{\delta S^{\mu}(\hat{i})\delta S_{\mu}^{}(\hat{i})}\phi[C_p^{}]
\nn
&=\frac{1}{2(D-p)\mathrm{Vol}[C_p^{}]}\int_{S^p}^{} d^p\xi\sqrt{h(\xi)}~\phi[C_p^{}]^*\frac{\delta^2}{\delta S^{\mu}(\xi)\delta S_{\mu}^{}(\xi)}\phi[C_p^{}]~
}
in the continuum-limit. 
In addition, the sum over all self-avoiding surfaces becomes the path-integral of embedding functions $\{X^\mu(\xi)\}_{\mu=1}^D$ as 
\aln{
a^D\sum_{C_p^{}\in \Gamma_p^{}}\quad \rightarrow \quad {\cal N}\int {\cal D}X~, 
\label{integral measure}
}
where ${\cal N}$ is a normalization factor (that should be determined appropriately) and ${\cal D}X$ is the path-integral measure induced by the diffeomorphism and reparametrization invariant norm
\aln{
||\delta X||^2\coloneq \int_{S^p}^{}d^p\xi \sqrt{h(\xi)} g_{\mu\nu}^{}(X(\xi))\delta X^{\mu}(\xi) \delta X^{\nu}(\xi)~.
}
It should be noted that Eq.~(\ref{integral measure}) contains the integral over  the center-of-mass position $\{x^\mu\}_{\mu=1}^D$ of $C_p^{}$ as 
\aln{
\int {\cal D}X=\int_{\Sigma_D^{}}\star 1\cdots ,
\label{path-integral measure}
} 
because a spacetime translation of a surface $C_p^{}$ results in another surface $C'_p$.  
This is the reason why we have included the factor $a^D$ in Eq.~(\ref{integral measure}). 

Expanding $(k_0^{}+\hat{H})^{-1}$ in powers of $\hat{H}-1$ and performing a field redefinition (without including the inverse gauge coupling $\beta$), we finally obtain the continuum-limit of Eq.~(\ref{loop functional theory}) as 
\aln{
&Z_{\text{Landau}}^{(p)}(\beta)\coloneq \int {\cal D}\phi {\cal D}\phi^* e^{-S_\mathrm{E}^{}}~,
\label{continuum partition function}
\\
S_{\mathrm{E}}^{}={\cal N}\int {\cal D}X &\left[U(\phi,\phi^*)-\frac{1}{\beta \mathrm{Vol}[C_p^{}]}\int_{S^p}^{} d^p\xi\sqrt{h(\xi)}~\phi[C_p^{}]^*\frac{\delta^2}{\delta S^{\mu}(\xi)\delta S_{\mu}^{}(\xi)}\phi[C_p^{}]
\right]
\nn
&+(\text{higher area-derivative terms})~,
\label{continuum action}
}
where 
\aln{U(\phi,\phi^*)\coloneq g^2 T_p^2 \phi^*\phi+V(\phi,\phi^*)
\label{potential U}
} 
with $T_p^{2}\propto a^{-2(p+1)}/(k_0^{}+1)$, which corresponds to a {\it bare} $p$-brane tension. 
%
%
%

%
In the continuum theory~(\ref{continuum partition function}), the $p$-form global transformation~(\ref{scalar p-form transformation}) becomes 
\aln{
\phi[C_p^{}]\quad \rightarrow\quad \begin{dcases}e^{i\theta\int_{C_p^{}} \Lambda_p^{}}\phi[C_p^{}] &  \text{for } G=\mathrm{U}(1)
\\
e^{\frac{i}{N}\int_{C_p^{}} \Lambda_p^{}}\phi[C_p^{}] &  \text{for } G=\mathbb{Z}_N^{}
\end{dcases}, 
}
where $\theta \in \mathbb{R}$ and $\Lambda_p^{}$ is a closed differential $p$-form satisfying $\int_{C_p^{}}\Lambda_p^{}\in 2\pi \mathbb{Z}$. 
In fact, one can check that the kinetic term in Eq.~(\ref{continuum action}) is invariant under these transformations due to  Eq.~(\ref{property of area derivative}) and the closeness of $\Lambda_p^{}$. 
%
%
%
For $G=\mathrm{U}(1)$, we can calculate the corresponding conserved current by the N\"{o}ther method as~\cite{Hidaka:2023gwh,Kawana:2024fsn,Kawana:2025vvf}
\aln{
J_{p+1}^{}(X)={\cal N}\int {\cal D}X\frac{i}{\mathrm{Vol}[C_p^{}](p+1)!}\int_{S^p} & d^p\xi \sqrt{h(\xi)}\left(\frac{\delta \phi^*[C_p^{}]}{\delta S^{[\mu_1^{}}(\xi)}\phi[C_p^{}]-\phi^*[C_p^{}]\frac{\delta \phi[C_p^{}]}{\delta S^{[\mu_1^{}}(\xi)}
\right)E_{\mu_2^{}\cdots \mu_{p+1}^{}]}(X(\xi))
\nn
&\times \frac{\delta^{(D)}(X-X(\xi))}{\sqrt{g(X(\xi))}}
dX^{\mu_1^{}}\wedge \cdots \wedge dX^{\mu_{p+1}^{}}~,
\label{p+1 form current}
}
which satisfies $d\star J_{p+1}^{}(X)=0$ when $\phi[C_p^{}]$ obeys the equation of motion. 
Then, the conserved charge is given by 
\aln{
Q_{p}^{}[C_{D-p-1}^{}]=\int_{C_{D-p-1}^{}}\star J_{p+1}^{}~,
}
where $C_{D-p-1}^{}$ is a $(D-p-1)$-dimensional closed subspace.  
It is a good exercise to check the transformation law 
\aln{
Q_p^{}[C_{D-p-1}^{}]\phi[C_p^{}]=\mathrm{Link}[C_{D-p-1}^{},C_p^{}]\phi[C_p^{}]~
}  
in the path-integral formulation~\cite{Hidaka:2023gwh}.

\subsection{Mass dimensions and upper critical dimension}\label{mass dimensions}

Readers who are primarily interested in the mean-field analysis can  skip this part and proceed directly to the next section. 

As in ordinary quantum field theory, mass dimensions of operators and coupling constants are important for dimensional analysis.     
%
In the following, we represent the mass dimension of a general quantity ${\cal O}$ by $[{\cal O}]$. 
In the continuum action~(\ref{continuum action}), we can see 
\aln{
\left[\frac{\delta}{\delta S^\mu(\xi)}\right]=\left[\frac{\delta}{\delta \sigma^{\mu_1^{}\cdots \mu_{p+1}^{}}(\xi)}\right]=p+1~,\quad [{\cal N}{\cal D}X]=-D~
}
by the definitions (\ref{def of area derivative}) and (\ref{integral measure}). 
Then, the mass dimension of the scalar field is found to be 
\aln{
[\phi[C_p^{}]]=\frac{D-2(p+1)}{2}~,
\label{mass dimension of string field}
} 
which is consistent with the ordinary mass dimension of scalar field $[\phi(X)]=(D-2)/2$ for $p=0$ and the previous results~\cite{Iqbal:2021rkn,Hidaka:2023gwh,Kawana:2024fsn} for $p=1$ as well.  

Once the mass dimension of the scalar field is known, mass dimensions of various coupling constants can be also determined. 
For example, focussing on local potential terms such as 
\aln{
U(\phi,\phi^*)=\mu \phi[C_p^{}]^*\phi[C_p^{}]+\lambda (\phi[C_p^{}]^* \phi[C_p^{}]))^2+\cdots ~,
}
we have
\aln{
[\mu]=2(p+1)~,\quad [\lambda]=4(p+1)-D~.
}
In the next section, we will see that $T_p^{}\coloneq \sqrt{|\mu|}$ corresponds to a world-volume tension of $p$-brane. 
The mass dimension of self-coupling $\lambda$ suggests that a naive upper-critical dimension of the Landau theory is $D_{\mathrm{c}}^{}=4(p+1)$, as predicted by  the Parisi's heuristic argument based on Hausdorff dimension of random surfaces~\cite{Parisi:1978mn} as follows. 

The dimension of $A\cap B$ for general two objects $A$ and $B$ is  
\aln{
d(A\cap B)=d(A)+d(B)-D~. 
} 
For example, for $D=3$ and $d(A)=d(B)=2$, we have $d(A\cap B)=2+2-3=1$, i.e., the intersection of two surfaces in $3$-dimensional space is a line.   
By applying this formula to two random surfaces $\mathrm{S}_A^{}$ and $\mathrm{S}_B^{}$, we obtain 
\aln{
d(\mathrm{S}_A^{}\cap \mathrm{S}_B^{})=2D_{\rm H}^{}-D~,
}
where $D_{\mathrm{H}}^{}$ is the Hausdorff dimension of random surfaces. 
This implies that for $D>2D_\mathrm{H}^{}$ there is essentially no intersection  between two surfaces, and critical behaviors of the system shall be described by the mean-field dynamics around the Gaussian fixed point.  
For $D<2D_\mathrm{H}^{}$, however, two random surfaces intersect in general, suggesting that their interactions become relevant for critical behaviors of the system. 
Thus, we can naively conclude that the upper critical dimension of random surfaces is $D_\mathrm{c}^{}=2D_\mathrm{H}^{}$ by this heuristic argument.  

Then, the remaining question is the determination of Hausdorff dimension $D_\mathrm{H}^{}$, and here come some subtleties;  
The definition of Hausdorff dimension is subject to ambiguities of defining a characteristic size of random surfaces. 
For example, the Hausdorff dimension of non-critical closed string (i.e., $p=1$ and $D<1$) has been examined in many literatures~\cite{Distler:1988jv,Kawai:1991qv,Kawai:1993cj,Ambjorn:1997jf,Ambjorn:2013lza}, yielding various predictions due to a variety of choices of   characteristic size.   
Let us here provide a reasonable   argument based on universality or central-limit theorem. 
Regardless of a microscopic theory of random closed $p$-dimensional surface $C_p^{}$, the average world-volume $\langle \mathrm{Vol}[M_{p+1}^{}]\rangle$ of $n$-step random $p$-dimensional surfaces would behave as~\cite{Kawana:2024qmz}
\aln{
\langle \mathrm{Vol}[M_{p+1}^{}]\rangle_n^{}\sim a^{p+1}n^{1/2}\quad \text{for } n\rightarrow \infty
} 
as long as the central-limit theorem holds. 
This then implies that a typical length $L\coloneq \langle \mathrm{Vol}[M_{p+1}^{}]\rangle_n^{\frac{1}{p+1}}$ and $n$ are related as 
\aln{n(L)\sim \left(\frac{L}{a}\right)^{2(p+1)}~,
}
leading to the Hausdorff dimension of random surfaces as $D_\mathrm{H}^{}=2(p+1)$. 
Combined with the above argument, the upper critical dimension can be estimated as
\aln{D_\mathrm{c}^{}=4(p+1)~,
}
which in fact coincides with the critical dimension determined by $[\lambda]=D-4(p+1)=0$.  
However, as already mentioned in Refs.~\cite{Iqbal:2021rkn,Kawana:2024qmz,Kawana:2025vvf}, this estimation is  too naive and the existence of explicit $\mathrm{U}(1)$ breaking terms and non-local interactions would provably alter critical behaviors and result in a different upper-critical dimension.  
More dedicated studies are necessary for its precise determination.

\section{Phases of higher gauge theory}\label{sec4}
In this section, we study phases of higher gauge theories in the Landau theory framework.    
See also Refs.~\cite{Iqbal:2021rkn,Hidaka:2023gwh,Kawana:2024fsn} for similar mean-field analyses.    

By taking the variation of the action~(\ref{continuum action}) with respect to $\phi[C_p^{}]$, we obtain the equation of motion as  
\aln{
\frac{1}{\mathrm{Vol}[C_p^{}]}\int_{S^p}^{} d^p\xi\sqrt{h(\xi)}\frac{\delta^2\phi[C_p^{}]}{\delta S^{\mu}(\xi)\delta S_{\mu}^{}(\xi)}-\frac{1}{g^2} \frac{\delta U(\phi,\phi^*)}{\delta \phi^*[C_p^{}]}=0~. 
\label{EOM of brane field}
}    
Let us investigate solutions of this equation in the strong and weak coupling limits respectively.    

\subsection{Strong coupling limit}
In the strong coupling limit $g\rightarrow \infty$, we can neglect the interaction terms in the potential~(\ref{potential U}) as 
\aln{
\frac{1}{g^2}U(\phi,\phi^*)=T_p^2\phi^*\phi+\frac{1}{g^2}V(\phi,\phi^*)\approx T_p^2\phi^*\phi\quad (\text{for }g\rightarrow \infty)
}
with $T_p^{2}>0$. 
In this case, Eq.~(\ref{EOM of brane field}) has a trivial solution $\langle \phi[C_p^{}]\rangle\approx 0$, which means that the system is in the unbroken phase of the $p$-form global symmetry.   
Compared to ordinary QFTs, however, this trivial solution should be interpreted as an asymptotic value of $\langle \phi[C_p^{}]\rangle$ in the large volume limit $\mathrm{Vol}[C_p^{}]\rightarrow \infty$ as follows.\footnote{
Although the invariance under spacetime translation forbids $\langle \phi[C_p^{}]\rangle$ to be dependent on the center-of-mass position of $C_p^{}$, it can still depend on the relative coordinates $\{Y^\mu(\xi)=X^\mu(\xi)-x^\mu\}_{\mu=1}^D$.   
}  
%
%

As a nontrivial solution of Eq.~(\ref{EOM of brane field}), we consider the following ansatz:
\aln{\phi[C_p^{}]=f(\mathrm{Vol}[M_{p+1}^{}])/\sqrt{2}~,\quad f(z)\in \mathbb{C}~, 
}
with a boundary condition $f(\infty)=0$, where $M_{p+1}^{}$ is the $(p+1)$-dimensional minimal surface surrounded by $C_p^{}$, i.e., $\partial M_{p+1}^{}=C_{p}^{}$. 
Putting this ansatz into Eq.~(\ref{EOM of brane field}), we obtain 
\aln{p(z)f''(z)+q(z)f'(z)-T_p^2f(z)\approx 0~,\quad z=\mathrm{Vol}[M_{p+1}^{}]~,
\label{simplified eom}
}
where 
\aln{p(z)&=\frac{1}{\mathrm{Vol}[C_p^{}]}\int_{S^p}d^p\xi\sqrt{h(\xi)} \left[\frac{1}{p!}(E_{p+1}^{})^{\mu \nu_1^{}\cdots \nu_{p}^{}}(E_{p}^{})_{\nu_1^{}\cdots \nu_{p}^{}}\right]^2~,
\\
q(z)&=\frac{1}{\mathrm{Vol}[C_p^{}]}\int_{S^p}d^p\xi\sqrt{h(\xi)} \frac{1}{p!}\frac{\delta [(E_{p+1}^{})^{\mu \nu_1^{}\cdots \nu_{p}^{}}(E_{p}^{})_{\nu_1^{}\cdots \nu_{p}^{}}]}{\delta S^{\mu}(\xi)}~.
}
For a concrete example, let us consider a spherical configuration $C_p^{}\simeq S^p$ specified by a fixed radius $r=R$ in the $(p+1)$-dimensional subspace with the polar coordinates $(r,\xi_1^{},\cdots,\xi_p^{})$.     
In this case, the projection tensors are explicitly calculated as  
\aln{(E_p^{})^{\nu_1^{}\cdots \nu_p^{}}&=\frac{\epsilon^{\nu_1^{}\cdots \nu_p^{}}}{R^{p}\sqrt{h(\xi)}}~,\quad \nu_i^{}\in \{\xi_1^{},\cdots,\xi_p^{}\}~,
\\
(E_{p+1}^{})^{\nu_1^{}\cdots \nu_{p+1}^{}}&=\frac{\epsilon^{\nu_1^{}\cdots \nu_{p+1}^{}}}{R^{p}\sqrt{h(\xi)}}~,\quad \nu_i^{}\in \{r,\xi_1^{},\cdots,\xi_p^{}\}~,
}
by which we obtain 
\aln{
p(z)=1~,\quad q(z)=0~.
}
Then, the equation of motion~(\ref{simplified eom}) takes a very simple form
\aln{
f''(z)-T_p^2f(z)\approx 0~,
\label{simplified EOM}
}
which allows the area-law solution 
\aln{
f(z)\approx c\times e^{-T_p^{}\mathrm{Vol}[M_{p+1}^{}]}~,
\label{Area law}
}
where $c$ is a constant that is determined by a boundary condition at $\mathrm{Vol}[C_p^{}]=0$ in principle.  
For more general configurations, $p(z)$ and $q(z)$ would become more complicated  functions of $z=\mathrm{Vol}[M_{p+1}^{}]$, but we can still argue that the large-volume behavior of the classical solution remains unchanged on dimensional ground.   
In fact, one can check $[p(z)]=0$ and $[q(z)]=p+1$, which implies that $p(z)\sim $ constant and $q(z)\sim z^{-1}$ for $z\rightarrow \infty$ as long as $C_p^{}$ is a non-singular surface characterized by a single length $R\sim \mathrm{Vol}[C_p^{}]^{\frac{1}{p}}\sim \mathrm{Vol}[M_{p+1}^{}]^{\frac{1}{p+1}}$.  
These functional behaviors of $p(z)$ and $q(z)$ lead to the same equation of motion as Eq.~(\ref{simplified EOM}) and result in the area-law in the large-volume limit.   

The area-law of the charged operator $\phi[C_p^{}]$ characterizes the unbroken $p$-form global symmetry and implies that the original $p$-form gauge theory is in the confined phase.  
Of course, this result had been anticipated from the strong coupling expansion in the gauge theory, but we have rederived it in the Landau theory framework. 
%

\subsection{Weak coupling limit}\label{sec:weak coupling}

Let us study the weak coupling limit $g\rightarrow 0$ next.  
In this case, the interaction terms in $U(\phi,\phi^*)/g^2$ cannot be neglected in general.    
%
Although the complete functional form of $U(\phi,\phi^*)$ is unknown, we can still argue that the $p$-form global symmetry must be spontaneously broken in this weak coupling limit by the following argument:     
By Eq.~(\ref{potential expansion}), the potential $U(\phi,\phi^*)$ is generally expanded as
\aln{
\frac{1}{g^2}U(\phi,\phi^*)=\left(T_p^2-\frac{1}{g^2}|\omega[C_p^{}]|^2\right)\phi^*\phi+(\text{higher-order terms})~
}  
in the vicinity of $\phi=0$, which shows that the quadratic term eventually becomes negative for a sufficiently small $|g|$, developing a nonzero vacuum expectation value (VEV) $\langle \phi[C_p^{}]\rangle=v/\sqrt{2}\neq 0$, regardless of the detailed structure of higher-order self interactions (as long as it is bounded below). 
%
Hence, the system is in the broken phase of the $p$-form global symmetry.  
Note that this mean-field result can be significantly altered by the proliferation of topological defects (i.e., monopoles for $p=D-2$), as we will see later.

In the Landau theory, the low-energy effective modes are identified by the phase modulations
\aln{
\phi[C_p^{}]=\frac{v}{\sqrt{2}}\exp\left(i\int_{C_p^{}}A_p^{}\right)~,
\label{phase modulation} 
}    
where $A_p^{}$ is a differential real-valued $p$-form field.  
By putting this into Eq.~(\ref{continuum action}), we can calculate the effective action of $A_p^{}$, which must respect the original $p$-form symmetry. 
For example, the effective action must be invariant under 
\aln{
A_p^{}\quad\rightarrow \quad \begin{dcases}
A_p^{}+\theta\Lambda_{p}^{}  & \text{for }G=\mathrm{U}(1)
\\
A_p^{}+\frac{1}{N}\Lambda_p^{} & \text{for }G=\mathbb{Z}_N^{}
\end{dcases}~,
\label{shift symmetry}
} 
where $\theta \in \mathbb{R}$ and $\Lambda_{p}^{}$ is a differential closed $p$-form satisfying $\int_{C_p^{}}\Lambda_p^{}\in 2\pi \mathbb{Z}$.  

Let us see each case below.       

\

\noindent{\bf $\mathrm{U}(1)$ HIGHER GAUGE THEORY}\\ 
In this case, the effective action in the absence of topological defects is the simple $p$-form Maxwell theory~\cite{Hidaka:2023gwh}
\aln{
S_\mathrm{eff}^{}[A_p^{}]=\frac{v^2}{2\beta}\int_{\Sigma_D^{}} F_{p+1}^{}\wedge \star F_{p+1}^{}~,\quad F_{p+1}^{}=dA_p^{}~,
\label{naive effective action}
}
where we have used the differential property~(\ref{property of area derivative}) of the area derivatives.    
See Appendix~\ref{app:effective theory} for the details.  
%
This theory describes a gapless higher-form field $A_p^{}$ with a gauge symmetry $A_p^{}\rightarrow A_p^{}+d\Lambda_{p-1}^{}$, where $\Lambda_{p-1}^{}$ is a differential $(p-1)$-form gauge parameter.    
%
%
In the subsection~\ref{sec:CMW}, however, we will see that the presence of topological defects renders $A_p^{}$ (or its dual field $\tilde{A}_{D-p-2}^{}$) massive and prevents the spontaneous breaking of compact $\mathrm{U}(1)$ higher-form symmetry for $p\geq D-2$.        
%

In the case of non-compact $\mathrm{U}(1)$ higher gauge theories, on the other hand, strong infra-red (IR) fluctuations of these gapless modes can spoil the spontaneous breaking of higher-form global symmetries~\cite{Lake:2018dqm}.      
%
In fact, they contribute to the expectation value of the Wilson-surface $W[C_p^{}]$ as~
\aln{
\langle W[C_p^{}]\rangle&=\left\langle e^{i\int_{C_p^{}}A_p^{}}\right\rangle=\exp\left(-\frac{g^2}{2}\int_{\Sigma_D^{}}\delta_{D-p}^{}(C_p^{})\wedge \star \hat{D}\delta_{D-p}^{}(C_p^{})\right)
\\
&=\exp\left(-\frac{\mathrm{Vol}[C_p^{}]}{2}\int \frac{d^{D-p}k}{(2\pi)^{D-p}}\frac{1}{k^2}\right)
\label{quantum VEV}
} 
where $\delta_{D-p}^{}$ is the Poincare dual-form of $C_p^{}$, i.e., 
\aln{\int_{\Sigma_D^{}}f_p^{}\wedge \delta_{D-p}^{}(C_p^{})=\int_{C_p^{}}f_p^{}
}
for ${}^\forall $ $p$-form $f_p^{}$, and $\hat{D}=d^\dagger d+\alpha^{-1}dd^\dagger$, where $\alpha$ is the gauge-fixing parameter.
The UV divergence in the exponent in Eq.~(\ref{quantum VEV}) can be always  removed by the redefinition of $W[C_p^{}]$. 
For $p\geq D-2$, however, we also encounter an IR divergence, which cannot be (simultaneously) removed by the field redefinition. 
As a result, it leads to $\langle W[C_p^{}]\rangle \rightarrow 0$, and the $p$-form global symmetry cannot be spontaneously broken.    
This is the Coleman-Mermin-Wagner theorem for non-compact $\mathrm{U}(1)$ higher-form global symmetries~\cite{Gaiotto:2014kfa,Lake:2018dqm}. 

\

\noindent{\bf FINITE HIGHER GAUGE THEORY}\\ 
When $G$ is a finite abelian group, the phase modulation~(\ref{phase modulation}) is no longer gapless because the scalar potential $U(\phi,\phi^*)$ produces a mass for $A_p^{}$.  
Regardless of a detailed functional form of $U(\phi,\phi^*)$, the effective potential $V(A_p^{})$ must respect the original discrete $p$-form (shift) symmetry. 
For example, when $G=\mathbb{Z}_N^{}$, it satisfies 
\aln{
V\left(A_p^{}+\frac{1}{N}\Lambda_p^{}\right)=V(A_p^{})~,
}
where $\Lambda_p^{}$ is a differential closed $p$-form with $\int_{C_p^{}}\Lambda_p^{}\in 2\pi \mathbb{Z}$. 
In the low-energy limit, we can use the Villain formula~\cite{Villain:1977} as 
\aln{
&\exp\left(-\int_{\Sigma_D^{}}V(A_p^{})\star 1\right)\approx \sum_{n\in \mathbb{Z}}\exp\left({\cal N}\int {\cal D}X\left\{1-\frac{1}{2}\left(N\oint_{C_p^{}} A_p^{}-2\pi n\right)^2\right\}
\right)
}
which can be further rewritten in a path-integral form by introducing two auxiliary higher-form fields as
\aln{
=&\int {\cal D}f_p^{}\int {\cal D}B_{D-p-1}^{}\exp\left({\cal N}\int {\cal D}X\left\{1-\frac{1}{2}\left(\oint_{C_p^{}} (NA_p^{}-f_p^{})\right)^2\right\}-\frac{i}{2\pi}\int_{\Sigma_D^{}}B_{D-p-1}^{}\wedge df_p^{} 
\right)~
}
with $\int_{C_{p}^{}}f_p^{}\in 2\pi \mathbb{Z}$. 
Then, by repeating similar calculations as Appendix~\ref{app:effective theory}, we obtain
\aln{
\int {\cal D}f_p^{}\int {\cal D}B_{D-p-1}^{}\exp\left(-\int_{\Sigma_D^{}}\frac{\lambda_2^{}}{2}\left(A_p^{}-\frac{f_p^{}}{N}\right)\wedge \star \left(A_p^{}-\frac{f_p^{}}{N}\right)-\frac{i}{2\pi}\int_{\Sigma_D^{}}B_{D-p-1}^{}\wedge df_p^{} 
\right)~,
} 
where $\lambda_2^{}$ is a coupling constant. 
We can eliminate $f_p^{}$ by the equation of motion as
\aln{
\frac{f_p^{}}{N}=A_p^{}+i\frac{N(-1)^{(D-p)(p+1)}}{2\pi \lambda_2^{}}\star dB_{D-p-1}^{}~,
} 
and the effective action becomes 
\aln{
S_{\rm eff}^{}&[A_p^{},B_{D-p-1}^{}]=\frac{v^2}{2\beta}\int_{\Sigma_D^{}}F_{p+1}^{}\wedge \star F_{p+1}^{}
\nn
+\int_{\Sigma_D^{}}&\left[\frac{N^2(-1)^{D-p}}{8\pi^2 \lambda_2^{}}dB_{D-p-1}^{}\wedge \star dB_{D-p-1}^{}-\frac{N^2(-1)^{(D-p)(p+1)}}{4\pi^2 \lambda_2^{}}d(B_{D-p-1}^{}\wedge \star dB_{D-p-1}^{})-i\frac{N}{2\pi}B_{D-p-1}\wedge dA_p^{}
\right]~.
}
The low-energy physics of the system is captured by neglecting all kinetic terms, resulting in  
\aln{
S_{\rm eff}^{}[A_p^{},B_{D-p-1}^{}]\approx -i\frac{N}{2\pi}
\int_{\Sigma_D^{}}B_{D-p-1}\wedge dA_p^{}~.
\label{topological field theory}
}
which is a $\mathrm{BF}$-type topological field theory. 
In addition to the original $\mathbb{Z}_N^{}$ $p$-form global symmetry~(\ref{shift symmetry}), Eq.~(\ref{topological field theory}) has an emergent $\mathbb{Z}_N^{}$ $(D-p-1)$-form global symmetry:
\aln{
B_{D-p-1}^{}\quad \rightarrow \quad B_{D-p-1}^{}+\frac{1}{N}\Lambda_{D-p-1}^{}~,\quad d\Lambda_{D-p-1}^{}=0~,\quad \int_{C_{D-p-1}}^{}\Lambda_{D-p-1}^{}\in 2\pi \mathbb{Z}~,
\label{emergent (D-p-1) symmetry}
}
where $\Lambda_{D-p-1}^{}$ is a $(D-p-1)$-form differential form and $C_{D-p-1}^{}$ is a $(D-p-1)$-dimensional closed subspace in $\Sigma_D^{}$. 
The charged objects are the Wilson-surfaces  
\aln{
W[C_p^{}]=\exp\left(i\int_{C_p^{}} A_p^{}\right)~,\quad U[C_{D-p-1}^{}]=\exp\left(i\int_{C_{D-p-1}^{}} B_{D-p-1}^{}\right)~. 
}
The topological theory (\ref{topological field theory}) exhibits topological order when the topology of spacial manifold $\Sigma_{D-1}^{}$ is nontrivial. 
See Refs.~\cite{Gomes:2023ahz,Bhardwaj:2023kri,Hidaka:2023gwh,Kawana:2024fsn} and references therein for more details about topological order. 
%

\subsection{Topological defects}\label{defect}

As in ordinary QFTs with $0$-form global symmetries, the equation of motion~(\ref{EOM of brane field}) in the Landau theory admits topologically nontrivial solutions, i.e., topological defects. 
In the following, we explicitly construct such a topological defect for $G=\mathrm{U}(1)$ and $\mathbb{Z}_N^{}$, respectively.  

\

\noindent {\bf COMPACT $\mathrm{U}(1)$ HIGER GAUGE THEORY}\\
In the previous subsection, we have seen that the effective theory of $A_p^{}$ is the $p$-form Maxwell theory~(\ref{naive effective action}) in the absence of topological defects. 
In addition to the original $\mathrm{U}(1)$ (electric) $p$-form global symmetry $d\star F_{p+1}^{}=0$, this effective theory has an emergent (magnetic) $\mathrm{U}(1)$ $(D-p-2)$-form symmetry $dF_{p+1}^{}=0$, under which the 't~Hooft-surface operator 
\aln{
U[M_{D-p-2}^{}]=e^{i\int_{M_{D-p-2}^{}}\tilde{A}_{D-p-2}^{}}
}
is charged, where $\tilde{A}_{D-p-2}^{}$ is the dual field of $A_p^{}$ and $M_{D-p-2}^{}$ is a $(D-p-2)$-dimensional topological defect coupled to $\tilde{A}_{D-p-2}^{}$.  
%
%
Below, we explicitly construct such a topological defect as a static solution of Eq.~(\ref{EOM of brane field}).   
This is a higher-dimensional version of the conventional global vortex solution for $\mathrm{U}(1)$ $0$-form global symmetry.

\begin{figure}
    \centering
    \includegraphics[scale=0.3]{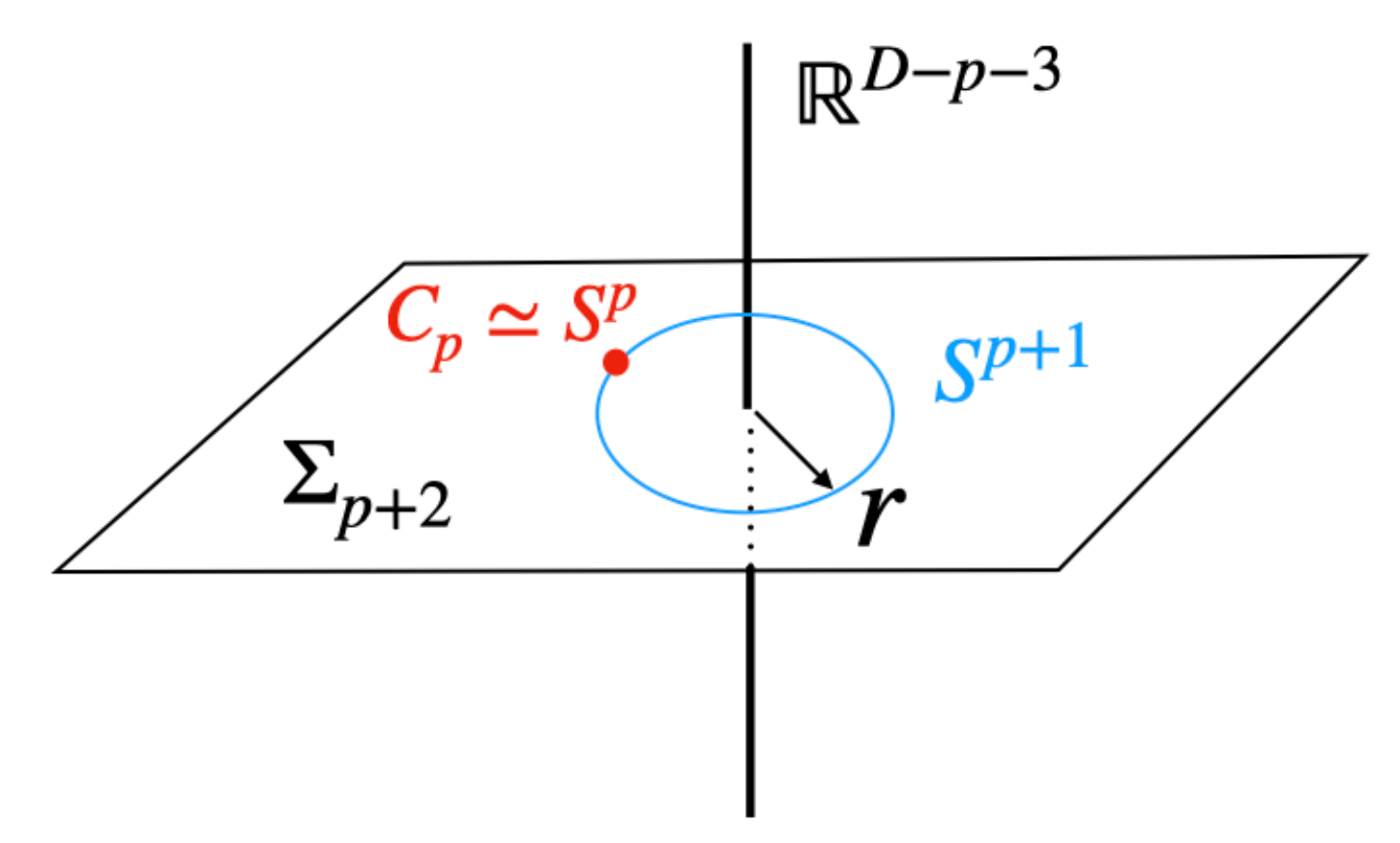}
    \caption{
   A topological defect in the compact $\mathrm{U}(1)$ $p$-form gauge theory.     
    The red point corresponds to the embedded $p$-brane $C_p^{}\simeq S^p$ that winds along $S^{p+1}$.  
    }
    \label{fig:vortex}
\end{figure}
\begin{figure}
    \centering
    \includegraphics[scale=0.3]{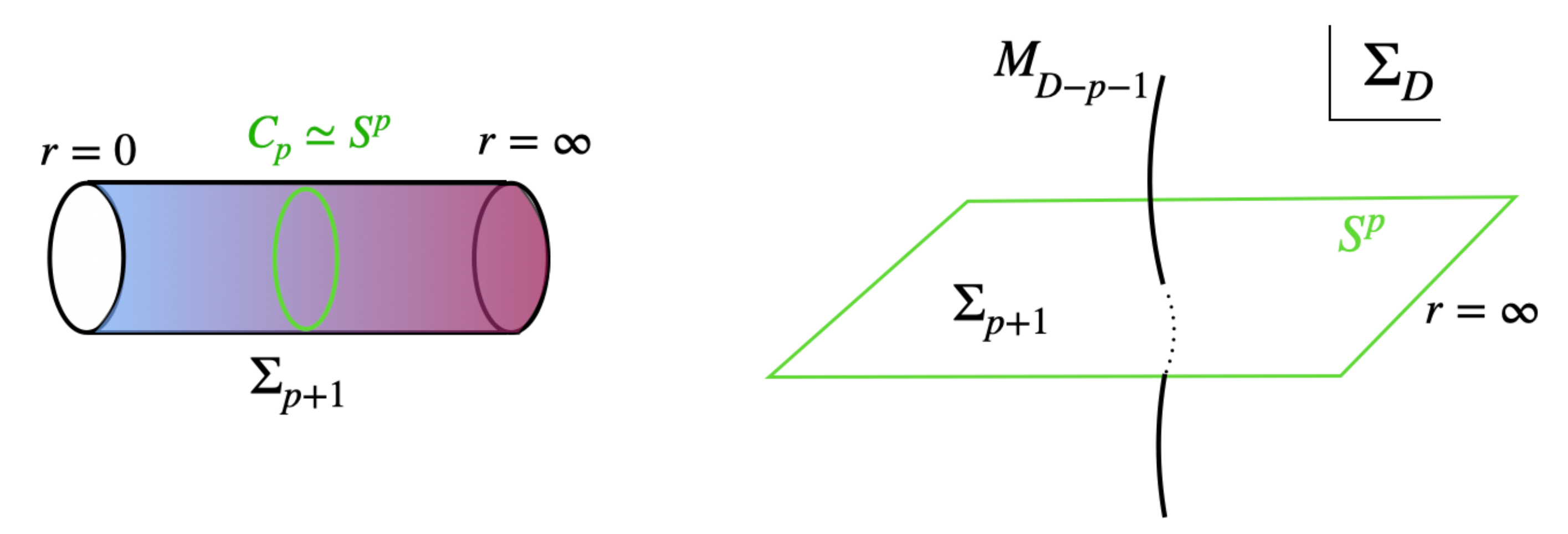}
    \caption{
    Left: Topologically nontrivial static configuration in $\mathrm{Z}_N^{}$ higher gauge theories.    
    The green circle represents the embedded surface $C_p^{}\simeq S^p$.   
    \\
    Right: Linking between the world-manifold $M^{}_{D-p-1}$ of topological defect and $S^{p}$. 
    }
    \label{fig:soliton}
\end{figure}

For simplicity, we consider the flat space $\Sigma_{D-1}^{}=\mathbb{R}^{D-1}$ and represent it as 
\aln{
\mathbb{R}^{D-1}=\mathbb{R}^{D-p-3}\times \Sigma_{p+2}^{}\coloneq \mathbb{R}^{D-p-3}\times S^{p+1}\times [0,\infty)~,
}
where the coordinate $r$ of $[0,\infty)$ corresponds to the radius of the $(p+2)$-dimensional subspace $\Sigma_{p+2}^{}$. 
See Fig.~\ref{fig:soliton} as an illustration.  
Besides, we represent the volume-form of the $q$-dimensional sphere $S^q$ as $\Omega_q^{}$, i.e.,
\aln{
\int_{S^q} \Omega_q^{}=\mathrm{Vol}[S^{q}]~.
}
For $q=0$, we define $\mathrm{Vol}[S^{0}]=1$.

We consider a spherical $p$-brane configuration $C_p^{}\simeq S^p$ embedded in $S^{p+1}$ and specified by $(r,\theta_1^{})$, where $\theta_1^{}$ is one of the polar coordinates $\theta_1^{}\in [0,\pi]$ of $S^{p+1}$.  
This is schematically shown by a red point in Fig.~\ref{fig:vortex}. 
This spherical configuration corresponds to reducing the (path-)integral measure of $C_p^{}$ as
\aln{
{\cal N}\int {\cal D}X\quad \rightarrow \quad 
\int_0^\infty dr r^{p+1}\int_0^{\pi} d\theta_1^{}(\sin\theta_1^{})^p
~. 
\label{minispace}
}  
Besides, the volume of $C_p^{}$ is expressed as 
\aln{
{\rm Vol}[C_p^{}]=(r\sin\theta_1^{})^p\int_{S^{p}}\Omega_p^{}
=(r\sin\theta_1^{})^p\times \mathrm{Vol}[S^{p}]~.
}
%
%
We then consider the following ansatz
\aln{
\phi_{}^{}[C_p^{}]=\frac{1}{\sqrt{2}}\left(\int_{C_{p}^{}}\chi_p^{}\right)
\exp\left(i\int_{C_p^{}}A_{p}^{\text{defect}}\right)~
\label{U(1) ansatz}
}
with 
\aln{
\chi_p^{}
=\chi(r)(\sin\theta_1^{})^{p}\Omega_p^{}~,\quad \chi(r)\in \mathbb{R}~,
}
and $F_{p+1}^{\text{defect}}:=dA_{p}^{\text{defect}}$ is the normalized volume-form on $S^{p+1}=C_p^{}\times [0,\pi]$ satisfying 
\aln{
\int_{S^{p+1}} F_{p+1}^{\text{defect}}=2\pi q~,\quad q\in \mathbb{Z}~.
\label{theta normalization}
}   
This normalization guarantees the single-valuedness of the scalar field $\phi[C_p^{}]$ in the volumeless limit as
%
\aln{\phi[C_p^{}]|_{\theta_1^{}=0}=\phi[C_p^{}]|_{\theta_1^{}=\pi}~.
}
%
Note that Eq.~(\ref{theta normalization}) is nothing but the Dirac quantization condition for the field strength of $A_p^{}$.  
%
\begin{figure}
    \centering
    \includegraphics[scale=0.7]{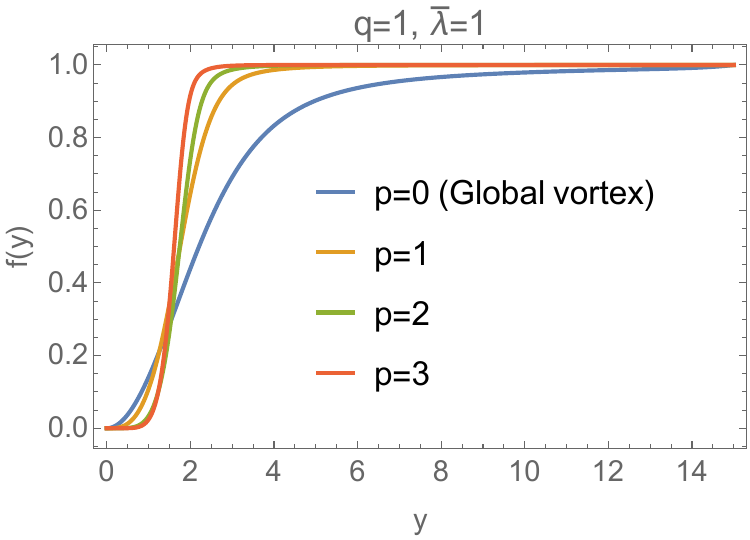}
    \caption{
    Field profiles of the topological defect for $G=\mathrm{U}(1)$. 
    Each color corresponds to a different choice of $p$.  
    }
    \label{fig:field profile for U(1)}
\end{figure}

Denoting the world-manifold of the defect as 
\aln{
M^{}_{D-p-2}\coloneq \mathbb{R}\times \mathbb{R}^{D-p-3}~,
}
we can also relate $F_{p+1}^{\text{defect}}$ to the Poincare dual-form of the world-manifold as follows:\footnote{
This expression of the Poincare dual is valid only for spherically symmetric  functions. 
Otherwise, we must include the delta functions for angular variables.     
}
\aln{\delta_{p+2}^{}(M^{}_{D-p-2})=\frac{1}{2\pi q}dF_{p+1}^{\text{defect}}~.
\label{Poincare dual and magnetic field}
}
%
In fact, one can check 
\aln{
\int_{\Sigma_{D}^{}}\delta_{p+2}^{}(M^{}_{D-p-2})\wedge \delta_{D-p-2}^{}(\Sigma_{p+2}^{})&=\int_{\Sigma_{p+2}}\delta_{p+2}^{}(M^{}_{D-p-2})=1~,
}
which represents the linking between $M_{D-p-2}^{}$ and $S_{}^{p+1}$. 
%

%
The area derivative on the ansatzs~(\ref{U(1) ansatz}) is evaluated as 
\aln{\frac{\delta \phi[C_p^{}]}{\delta S^{\mu}}&=\frac{1}{p!}(E_p^{})^{\nu_1^{}\cdots \nu_p^{}}\left(d\chi_p^{}+i\left(\int_{C_p^{}}\chi_p^{}\right)dA_p^{\text{defect}}\right)_{\mu\nu_1^{}\cdots \nu_p^{}}^{}e^{i\int_{C_p^{}}A_p^{\text{defect}}}
\nn
&=\frac{1}{r^p}\left[\delta_\mu^r \frac{d\chi}{dr}+\delta_\mu^{\theta_1^{}}\left(\frac{p}{\sin\theta_1^{}}+i\frac{2\pi q\mathrm{Vol}[S^{p}](\sin\theta_1^{})^p}{\mathrm{Vol}[S^{p+1}]}\right)\chi \right]e^{i\int_{C_p^{}}A_p^{\text{defect}}}
~,
}
which leads to the effective action (energy) of $\chi(r)$ as
\aln{
&\int_0^{\infty}dr\int_0^{\pi}d\theta_1^{}\frac{(\sin\theta_1^{})^p}{r^{p-1}}\left[\frac{1}{2}\left(\frac{d\chi}{dr}\right)^2+\frac{1}{2r^2}\left\{\frac{p^2}{(\sin\theta_1^{})^2}+\left(\frac{2\pi q\mathrm{Vol}[S^{p}](\sin\theta_1^{})^p}{\mathrm{Vol}[S^{p+1}]}\right)^2\right\}\chi^2+r^{2p}U(\phi)
\right]
\nn
&\propto \int_0^{\infty}dr\frac{1}{r^{p-1}}\left[\frac{1}{2}\left(\frac{d\chi}{dr}\right)^2+\frac{c_2^{}}{2r^2}\chi^2+r^{2p} \tilde{U}(\chi)
\right]~,
\label{vortex action}
}
where $c_2^{}$ is a numerical constant specified by $p$ and $q$, and $\tilde{U}(\chi)$ is the effective potential of $\chi$ after integrating out $\theta_1^{}$. 
Equation~(\ref{vortex action}) reduces to the conventional vortex energy for $p=0$. 
By introducing dimensionless quantities  
\aln{
f=\frac{\chi}{v}~,\quad y=vr~,\quad \overline{U}(f)=\frac{1}{v^4}\tilde{U}(\chi)~, 
} 
we obtain the dimensionless equation of motion: 
\aln{
\frac{1}{y^{p+1}}\frac{d}{dy}\left(\frac{1}{y^{p-1}}\frac{df}{dy}\right)-\frac{c_2^{}}{y^{2(p+1)}}f-\frac{\delta \overline{U}(f)}{\delta f}=0
\label{dimensionless eom}
}
with boundary conditions $f(0)=0$ and $f(\infty)=1$. 
For simplicity, we consider a local potential 
\aln{\overline{U}(f)=\frac{\lambda}{4}(f^2-1)^2~,\quad \lambda>0~,
}
for numerical calculations. 
In Fig.~\ref{fig:field profile for U(1)}, we show numerical plots of  $f(y)$ for $p=0$ (blue), $p=1$ (orange), $p=2$ (green), and $p=3$ (red) with $\overline{\lambda}=1$ and $q=1$. 
The blue one corresponds to the conventional global vortex profile.  
One can see that the variation of $p$ corresponds to the change of the slope of $f(y)$ in the vicinity of the origin, as expected from the equation of motion~(\ref{dimensionless eom}).   

\

\noindent {\bf FINITE HIGHER GAUGE THEORIES}\\
Next let us consider the discrete case $G=\mathbb{Z}_N^{}$. 
The emergent $\mathbb{Z}_N^{}$ $(D-p-1)$-form symmetry~(\ref{emergent (D-p-1) symmetry}) in the effective theory implies that there must exist a $(D-p-1)$-dimensional defect $M_{D-p-1}^{}$ coupled to the dual field $B_{D-p-1}^{}$, and we will construct it explicitly below.  
%

%
In this case, we represent the space as 
\aln{
\mathbb{R}^{D-1}=\mathbb{R}^{D-p-2}\times \Sigma_{p+1}^{}\coloneq \mathbb{R}^{D-p-2}\times S^{p}\times [0,\infty)~,
}
where $r\in [0,\infty)$ denotes the radius of $S^p$.  
%
%
Note that the boundary of $\Sigma_{p+1}^{}$ is  $\partial \Sigma_{p+1}^{}= \Sigma_{p+1}^{}|_{r=\infty}^{}\simeq S^{p}$.  
%
%
A $p$-dimensional surface $C_p^{}$ is embedded as $C_p^{}\simeq S^p$ in $\Sigma_{p+1}^{}$, as depicted by a green circle in the left panel in Fig.~\ref{fig:soliton}.  
This corresponds to the reduction of the path-integral measure as 
\aln{
{\cal N}\int {\cal D}X \quad \rightarrow \quad {\cal N}\int_0^\infty dr r^{p}
~.
\label{minispace}
} 
We then consider the following ansatz:
\aln{
\phi[C_p^{}]=\frac{v}{\sqrt{2}}\exp\left(i\int_{S^p}A_p^{\text{defect}}\right)~,
\quad A_{p}^{\text{defect}}=f(r)\frac{\Omega_{p}^{}}{\mathrm{Vol}[S^p]}~,
\label{ansatz}
}
with the boundary conditions
\aln{
f(0)=0~,\quad f(\infty)=\frac{2\pi}{N}n~,\quad n\in \mathbb{Z}~. 
}
Namely, Eq.~(\ref{ansatz}) represents a topological configuration interpolating two degenerate minima.  
The corresponding topological charge is 
\aln{
Q_{D-p-1}^{}=\frac{N}{2\pi}\int_{\Sigma_{p+1}^{}}dA_p^{\text{defect}}=\frac{N}{2\pi}\int_{S^p}A_p^{\text{defect}}\bigg|_{r=\infty}^{}=n\in \mathbb{Z}~,
}
where we have used the Stokes theorem in the second equality.

Denoting the world-manifold of the defect as 
\aln{
M^{}_{D-p-1}\coloneq \mathbb{R}\times \mathbb{R}^{D-p-2}~,
}
we can relate $A_p^{\text{defect}}$ to the Poincare dual of $M^{}_{D-p-1}$ as  
\aln{
\delta_{p+1}^{}(M^{}_{D-p-1})=\frac{N}{2\pi Q_{D-p-1}^{}}dA_p^{\text{defect}}~.\label{world-volume form}
}
In fact, we can check 
\aln{
\int_{\Sigma_{D}^{}}\delta_{p+1}(M^{}_{D-p-1})\wedge \delta_{D-p-1}^{}(\Sigma_{p+1}^{})=\int_{\Sigma_{p+1}}\delta_{p+1}^{}(M^{}_{D-p-1})=\frac{N}{2\pi Q_{D-p-1}^{}}\int_{\Sigma_{p+1}}dA^{\text{defect}}_{p}=1~,
}
which represents the linking between $M_{D-p-1}^{}$ and $C_p^{}\simeq S_{}^{p}$, as illustrated in the right panel in Fig.~\ref{fig:soliton}.  

The functional derivative on the ansatz~(\ref{ansatz}) is evaluated as 
\aln{
\frac{\delta \phi[C_p^{}]}{\delta S^\mu(\xi)}=
i \frac{1}{p!}E^{\nu_1^{}\cdots \nu_p^{}}(dA_p^{\text{defect}})_{\mu \nu_1^{}\cdots \nu_p^{}}\phi[C_p^{}]=-i\frac{1}{\mathrm{Vol}[S^{p}]r^p}\frac{df}{dr}\delta_\mu^{r}\phi[C_p^{}]~,
}
by which we obtain the effective action (energy) of $f(r)$ as 
\aln{
\int_0^\infty drr^p\left[\frac{v^2}{2\mathrm{Vol}[S^p]^2}\left(\frac{1}{r^p}\frac{df}{dr}\right)^2+U(f)\right]
\propto v^{p+3}\int_0^\infty dyy^p\left[\frac{1}{2}\left(\frac{1}{y^p}\frac{df}{dy}\right)^2+\overline{U}(f)\right]~,
\label{defect effective action}
}
where the dimensionless radius $y\coloneq (\mathrm{Vol}[S^p])^{\frac{1}{p+1}}vr$ and potential $\overline{U}\coloneqq \mathrm{Vol}[S^p]^2 U/v^{2p+4}$ are introduced.  
By taking the variation of Eq.~(\ref{defect effective action}) with respect to $f(y)$, we obtain the dimensionless equation of motion:\footnote{This can also  be derived from the general equation of motion~(\ref{EOM of brane field}) by putting the ansatz~(\ref{ansatz}) and using~(\ref{property of area derivative}).
}
\aln{
\frac{1}{y^p}\frac{d}{dy}\left(\frac{1}{y^p}\frac{df}{dy}\right)-\frac{\delta \overline{U}}{\delta f}=0~.
\label{defect equation}
}
In general, the periodic potential $\overline{U}(f)=\overline{U}\left(f+\frac{2\pi}{N}\right)$ is expressed by a Fourier series of $\cos(N k f)$ and $\sin(Nk f)$ with $k\in \mathbb{N}$.  
For a concrete example, let us consider the simplest one   
\aln{
\overline{U}(f)=\overline{U}_A^{}(f)\coloneq \overline{\lambda}\left(1-\cos\left(N f\right)\right)~,
\label{cosine potential}
} 
where $\overline{\lambda}$ is a dimensionless coupling. 
This potential appears from the first term in Eq.~(\ref{determinant terms}) under the ansatz~(\ref{ansatz}).  
In this case, $N$ dependence is essentially irrelevant because it can be eliminated in the equation of motion~(\ref{defect equation}) by $Nf\rightarrow f$ and $N^{\frac{1}{p}}y\rightarrow y$.  
Thus, it suffices to calculate the profile $f(y)$ for $N=1$. 
\begin{figure}
    \centering
    \includegraphics[scale=0.65]{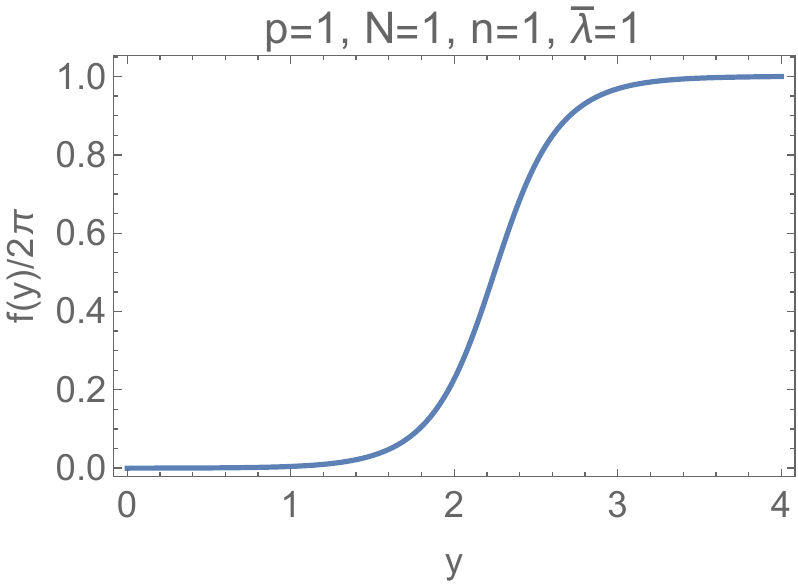}
    \caption{
    Field profile of the topological defect for $G=\mathbb{Z}_N^{}$.  
    }
    \label{fig:profile}
\end{figure}
\noindent 
In Fig.~\ref{fig:profile}, we show the numerical plot of $f(y)/2\pi$ for $N=1$, $p=1$, $n=1$ and $\overline{\lambda}=1$.      
%
%
%

We can also examine the low-energy effective theory in the presence of    topological defect.  
Instead of Eq.~(\ref{phase modulation}), the phase modulation $A_p^{}$ is introduced around the defect solution as 
\aln{
\phi[C_p^{}]=\frac{v}{\sqrt{2}}\exp\left(i\int_{C_p^{}} (A_p^{}+A_p^{\text{defect}})\right)~, 
} 
and the low-energy effective action is obtained by replacing $A_p^{}$ in Eq.~(\ref{topological field theory}) by $A_p^{}+A_p^{\text{defect}}$ as 
\aln{
S_{\rm eff}^{}[A_p^{},B_{D-p-1}^{}]\approx i\frac{N}{2\pi}\int_{\Sigma_{D}^{}}B_{D-p-1}^{}\wedge dA_p^{}+iQ_{D-p-1}^{}\int_{M_{D-p-1}^{}}B_{D-p-1}^{}~,
\label{effective theory in discrete}
}
where we have used Eq.~(\ref{world-volume form}). 
The second term corresponds to the interaction between the dual field $B_{D-p-1}^{}$ and the topological defect, as expected.  

\subsection{Coleman-Mermin-Wagner Theorem}\label{sec:CMW}

We discuss the Coleman-Mermin-Wagner theorem~\cite{Gaiotto:2014kfa,Lake:2018dqm} for higher-form global symmetries in the present Landau theory framework.  

\

\noindent {\bf COMPACT $\mathrm{U}(1)$ HIGHER GAUGE THEORY}\\
In subsection~\ref{sec:weak coupling}, we saw that non-compact $\mathrm{U}(1)$ $p$-form global symmetry cannot be spontaneously broken for $p\geq D-2$ due to the IR divergence.      
The same conclusion holds in compact $\mathrm{U}(1)$ case as well with a  different underlying mechanism; The gas of topological defects renders $A_p^{}$ massive, and thereby preventing spontaneous symmetry breaking. 
In the presence of a single topological defect, low-energy fluctuations are introduced by 
\aln{
\phi[C_p^{}]=\frac{v}{\sqrt{2}}\exp\left(i\int_{C_p^{}}\left[A_p^{\text{defect}}+A_p^{}\right]\right)~,
}
where we have taken the thin limit of the defect for simplicity, i.e., $\int_{C_p^{}}\chi_p^{}\rightarrow v$.  
By putting this into the action~(\ref{continuum action}) and performing similar calculations to Eq.~(\ref{naive effective action}), we have
\aln{
S_{\mathrm{eff}}^{}[A_p^{}]=\frac{v^2}{2\beta}\int_{\Sigma_D^{}} (F_{p+1}^{}+F_{p+1}^\text{defect})\wedge \star (F_{p+1}^{}+F_{p+1}^\text{defect})~, 
\label{effective action in the presence of monopoles}
} 
where $F_{p+1}^{\text{defect}}$ corresponds to the magnetic field sourced by the defect and satisfies Eq.~(\ref{Poincare dual and magnetic field}). 
Since Eq.~(\ref{effective action in the presence of monopoles}) is a quadratic form, we can dualize it by introducing a $(D-p-1)$-form $\tilde{F}_{D-p-1}^{}$ as 
\aln{
\frac{\beta}{2v^2(2\pi)^2}\int_{\Sigma_D^{}}\tilde{F}_{D-p-1}^{}\wedge \star \tilde{F}_{D-p-1}^{}+\frac{i}{2\pi}\int_{\Sigma_D^{}}\tilde{F}_{D-p-1}^{}\wedge (F_{p+1}^{}+F_{p+1}^{\text{defect}})~,
\label{dual field transformation}
}
and the path-integral of the original field $A_p^{}$ leads to the flat condition $d\tilde{F}_{D-p-1}^{}=0$, that is, $\tilde{F}_{D-p-1}^{}=d\tilde{A}_{D-p-2}^{}$, where $\tilde{A}_{D-p-2}^{}$ is interpreted as a dual field of $A_p^{}$. 
Using Eq.~(\ref{Poincare dual and magnetic field}) and performing a partial integration, we obtain the dual effective theory in the presence of a single topological defect $M_{D-p-2}^{}$ as 
\aln{
&\int {\cal D}\tilde{A}_{D-p-2}^{}~e^{-S_{\rm eff}^{\rm dual}[\tilde{A}_{D-p-2}^{}]}~,
\label{dual effective theory}
\\
S_{\rm eff}^{\rm dual}[\tilde{A}_{D-p-2}^{}]=&\frac{\beta}{2v^2(2\pi)^2}\int_{\Sigma_D^{}} \tilde{F}_{D-p-1}^{}\wedge \star \tilde{F}_{D-p-1}^{}+iq\int_{M_{D-p-2}^{}}\tilde{A}_{D-p-2}^{}~,
\label{dual effective action}
}
%
where the second term in Eq.~(\ref{dual effective action}) is the interaction between the topological defect $M_{D-p-2}^{}$ and the dual field $\tilde{A}_{D-p-2}^{}$.  
%
%
%
%
For $D-p-2>0$, $M_{D-p-2}^{}$ is an extended object, and its proliferation is energetically disfavored. 

On the other hand, when $D-p-2=0$, $M_{D-p-2}^{}$ is a point-like object, i.e., monopole, and the summation over monopoles and their locations can lead to an effective potential of the dual scalar field $\varphi\coloneq \tilde{A}_{D-p-2}^{}$. 
In fact, the monopole summation gives  
\footnote{
Here, we assume that monopoles with charges $n=\pm 1$ contribute to the partition function because they are quite heavy in the weak coupling limit. 
}
\aln{
\sum_{N=0}^{\infty}\frac{(\xi/2)^N}{N!}\left(\prod_{k=1}^N\sum_{q_k^{}\in \pm 1}\int d^Dx_k^{}\right)e^{iq_k^{}\varphi(x_i^{})}=&\sum_{N=0}^{\infty}\frac{\xi^N}{N!}\left(\int d^Dx~2\cos(\varphi(x))\right)^N 
\nn
=&\exp\left[2\xi \int_{\Sigma_D^{}}\cos(\varphi(x))\star 1\right]~,
}
where $\xi$ is the fugacity of a single monopole.    
We then obtain the following effective theory in the weak coupling limit:
\aln{
\int {\cal D}\varphi \exp\left(-\int_{\Sigma_D^{}}\left[\frac{\beta^2}{2v^2(2\pi)^2}d\varphi\wedge \star d\varphi-\xi\cos(\varphi)\star 1 
\right]\right)~,
\label{monopole gas}
}    
which is a massive scalar theory and indicates that the $\mathrm{U}(1)$ $p$-form global symmetry must be preserved.  
In fact, we can explicitly check it by calculating the VEV $\langle \phi[C_p^{}]\rangle$ in  the effective theory~(\ref{monopole gas}). 
The presence of $\phi[C_p^{}]$ corresponds to adding 
\aln{
i\int_{\Sigma_D^{}}\delta_{2}^{}(C_p^{})\wedge A_p^{}~
}
to Eq.~(\ref{dual field transformation}), which then leads to the constraint $d\tilde{F}_{1}^{}=2\pi \delta_{2}^{}(C_p^{})$ by integrating out $A_p^{}$. 
Namely, we have
\aln{\langle \phi[C_p^{}]\rangle \approx \frac{1}{Z} \int {\cal D}\varphi \exp&\left(-\int_{\Sigma_D^{}}\left[\frac{\beta^2}{2v^2(2\pi)^2}d\varphi\wedge \star d\varphi-\xi \cos (\varphi)\star 1\right]\right)\times \delta(d\tilde{F}_{1}^{}-2\pi \delta_{2}^{}(C_p^{}))~,
\label{VEV in monopole gas}
}
where the scalar field $\varphi$ is non-singular on $\Sigma_D^{}\textbackslash C_p^{}$, and its field strength $\tilde{F}_1^{}=d\varphi$ satisfies 
\aln{
\int_{\Sigma_D^{}}d\tilde{F}_{1}^{}\wedge \delta_{D-2}(M_{2}^{})=\int_{\Sigma_D^{}}\tilde{F}_{1}^{}\wedge \delta_{D-1}(C_{1}^{})=2\pi~
\quad \therefore \quad \int_{C_1^{}}d\varphi=2\pi
}
for any loop $C_{1}^{}$ that links $C_p^{}$.  
Here, we have used the Stokes theorem in the first equality.  
To understand the qualitative behavior of Eq.~(\ref{VEV in monopole gas}) in the large volume limit, we do not need detailed information of an analytic saddle.   
\begin{figure}[t!]
\vspace{-2cm}
 \centering
 \includegraphics[scale=0.5]{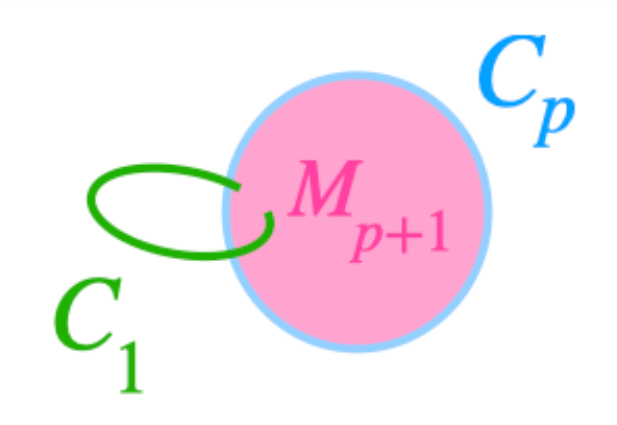}
\caption{
Winding of the dual scalar field $\varphi$ along a loop $C_1^{}$ that links $C_p^{}$. 
 }
 \label{fig:monopole}
\end{figure}
%
See Fig.~\ref{fig:monopole} for example, where the winding between $C_p^{}$ and $C_1^{}$ is illustrated in a $D=(p+2)$-dimensional space. 
Since the classical field $\varphi$ satisfies $d\star d\varphi-\xi \sin(\varphi)\star 1=0$ on $\Sigma_D^{}\textbackslash C_p^{}$, this winding is localized on the intersecting point between $C_p^{}$ and $C_1^{}$, that is, the minimal surface $M_{p+1}^{}$ bounded by $C_p^{}$, costing an energy $\sigma\propto \xi $ per unit volume on $M_{p+1}^{}$.    
%
%
As a result, the saddle-point Euclidean action is qualitatively given by    
\aln{S_{\text{saddle}}^{}\sim \sigma \mathrm{Vol}[M_{p+1}^{}]~,
}
which then corresponds to the area-law 
\aln{\langle \phi[C_p^{}]\rangle \sim e^{-\sigma \mathrm{Vol}[M_{p+1}^{}]}~.
}   
This implies that compact $\mathrm{U}(1)$ $p$-form global symmetry cannot be spontaneously broken for $p\geq D-2$ as in non-compact $\mathrm{U}(1)$ case.    
This is the higher-form version of the Polyakov’s argument for the confinement of compact $\mathrm{U}(1)$ gauge theory in three dimensions~\cite{Polyakov:1975rs,Polyakov:1976fu,Polyakov:1980ca,Panero:2005iu}. 

\

\noindent {\bf FINITE HIGHER GAUGE THEORY}\\
The derivation of the Coleman-Mermin-Wagner theorem for finite higher gauge theories is completely parallel to the above compact $\mathrm{U}(1)$ case just by replacing the dimensionality of topological defect from $(D-p-2)$ to $(D-p-1)$. 
%
In fact, we have already derived the low-energy effective theory~(\ref{effective theory in discrete}) in the presence of a single topological defect, and it is equivalent to Eq.~(\ref{dual effective action}) by identifying $B_{D-p-1}^{}$ with $\tilde{A}_{D-p-2}^{}$ and including the kinetic term of $B_{D-p-1}^{}$. 
%
Thus, repeating the same discussion, we conclude that discrete $p$-form global symmetry is not spontaneously broken for $p\geq D-1$. 
%

\section{Kramers-Wannier Duality}\label{sec5}
We discuss Kramers-Wannier (KW) duality in lattice higher gauge theories and its implication for infrared (IR) duality in continuum  Landau field theories.   

\subsection{Duality in higher gauge theories}\label{sec:duality}  
In this subsection, we derive the KW duality between the $p$-form lattice higher gauge theory and the $(D-p-2)$-form {\it gauged} higher gauge theory in our lattice setup.  
The meaning of {\it gauged} will be clarified soon below. 
For an earlier derivation, see also Ref.~\cite{Itzykson_Drouffe_1989}.
Readers familiar with this subject can skip this subsection.  

In the lattice setup introduced in subsection~\ref{Higher gauge theory on the lattice}, we define a {\it $p$-chain} by the abstract sum 
\aln{
\sum_{L_p^{}}\theta_{L_p^{}}L_p^{}~,\quad \theta_{L_p^{}}^{}\coloneq i^{-1}\log (U_{L_p^{}})~,
}
where the summation runs over all positive $p$-links. 
The boundary and coboundary operators are defined by 
\aln{
\Delta L_p^{}\coloneq \sum_{L_{p-1}^{}}\mathrm{Inc}(L_p^{},L_{p-1}^{})L_{p-1}^{}~,\quad  \nabla L_p^{}\coloneq \sum_{L_{p+1}^{}}\mathrm{Inc}(L_{p+1}^{},L_{p}^{})L_{p+1}^{}~,
} 
where $\mathrm{Inc}(L_p^{},L_{p-1}^{})$ is the incident number defined by 
\aln{
\mathrm{Inc}(L_p^{},L_{p-1}^{})=\begin{dcases}
0 & \text{if}~L_{p-1}^{}\not\subset \partial L_{p}^{}
\\
1 & \text{if}~L_{p-1}^{}\subset \partial L_{p}^{} \text{ and is positively oriented}
\\
-1 & \text{if}~L_{p-1}^{}\subset \partial L_{p}^{} \text{ and is negatively   oriented}
\end{dcases}. 
}
Then, a general gauge-invariant theory can be written as\footnote{
If the gauge invariance is not imposed, we can also add a potential such as $\sum_{L_p^{}}V(\theta_{L_p^{}}^{})$. 
}
\aln{
Z_G^{(p)}(\beta)&=\int [dU]e^{-S^{(p)}}~,
\quad  S^{(p)}=\beta \sum_{L_{p+1}^{}}F\left(\sum_{L_p^{}}{\rm Inc}(L_{p+1}^{},L_{p}^{})\theta_{L_p^{}}^{}\right)~,\label{general higher gauge theory}
}
where $F(\theta)$ is an arbitrary real function and the path-integral measure is defined by Eq.~(\ref{measure}). 
For example, the Wilson action~(\ref{lattice Ising model}) corresponds to $F(\theta)=\beta(1-\cos \theta)$. 
Moreover, the subscript $G$ in the partition function denotes the $p$-form global symmetry $G$ in the theory.    
The integrand in Eq.~(\ref{general higher gauge theory}) can be expressed  as a Fourier transform as 
\aln{
e^{-S^{(p)}}&=\prod_{L_{p+1}^{}}e^{-\beta F\left(\sum_{L_p^{}}{\rm Inc}(L_{p+1}^{},L_{p}^{})\theta_{L_p^{}}^{}\right)}
\nn
&=\prod_{L_{p+1}^{}}\left\{\sum_{\alpha_{L_{p+1}^{}}\in \widehat{G}} f(\alpha_{L_{p+1}^{}}^{})\chi_{\alpha_{L_{p+1}^{}}}^{}\left(\sum_{L_p^{}}{\rm Inc}(L_{p+1}^{},L_{p}^{})\theta_{L_p^{}}^{}\right)\right\}~,
\label{Fourier integrand}
}
where $\alpha_{L_{p+1}^{}}^{}\in \widehat{G}$ is a dual group element attached to $L_{p+1}^{}$, $\chi_\alpha^{}(\theta)=e^{i\alpha \theta}$ is the character of the representation $\alpha\in \widehat{G}$, and 
\aln{
f(\alpha)=\begin{dcases}\frac{1}{|G|}\sum_{\theta\in G}\chi_\alpha^{*}(\theta)e^{-\beta F(\theta)} & \text{for a finite abelian group $G$} 
\\
\frac{1}{2\pi}\int_{-\pi}^{\pi}d\theta \chi_\alpha^{*}(\theta)e^{-\beta F(\theta)} & \text{for $G=\mathrm{U}(1)$}
\end{dcases}~.
}
Using $\chi_\alpha^{}(\theta)\chi_\beta^{}(\theta)=\chi_{\alpha+\beta}^{}(\theta)$, we can group all terms depending on $\theta_{L_{p}}^{}$ in Eq.~(\ref{Fourier integrand}) as
\aln{
\prod_{L_{p+1}^{}} \chi_{\alpha_{L_{p+1}^{}}}^{}\left(\mathrm{Inc}(L_{p+1}^{},L_p^{})\theta_{L_p^{}}^{}\right)
=\chi_{\sum_{L_{p+1}^{}}\mathrm{Inc}(L_{p+1}^{},L_p^{})\alpha_{L_{p+1}}^{}}(\theta_{L_{p}^{}}^{})~,
} 
and the summation over $\theta_{L_p^{}}^{}$ provides  
\aln{
\frac{1}{|G|}&\sum_{\theta_{L_p^{}}\in G}\chi_{\sum_{L_{p+1}^{}}\mathrm{Inc}(L_{p+1}^{},L_p^{})\alpha_{L_{p+1}^{}}^{}}(\theta_{L_{p}^{}}^{})
=\delta_{0,\sum_{L_{p+1}^{}}\mathrm{Inc}(L_{p+1}^{},L_p^{})\alpha_{L_{p+1}^{}}}^{}~.
}
The same result holds for $G=\mathrm{U}(1)$.
Now, the partition function~(\ref{general higher gauge theory}) is written as 
\aln{
Z_G^{(p)}(\beta)=
\int [d\widehat{U}]e^{-\widehat{S}^{}}\times \left(\prod_{L_p^{}}\delta_{0,\sum_{L_{p+1}^{}}\mathrm{Inc}(L_{p+1}^{},L_p^{})\alpha_{L_{p+1}^{}}}^{}\right)~, 
\label{dual expression}
}
where 
\aln{
\widehat{S}^{}=-\sum_{L_{p+1}^{}}\log f(\alpha_{L_{p+1}^{}}^{})~,
\quad \int [d\widehat{U}]=\prod_{L_{p+1}^{}}\sum_{\alpha_{L_{p+1}^{}}^{}\in \widehat{G}}~.
}
Equation~(\ref{dual expression}) is similar to the original theory~(\ref{general higher gauge theory}), but each link variable is an element of $\widehat{G}$ and attached to $L_{p+1}^{}$ instead of $L_p^{}$.   
We can further express it in a dual form by regarding that these link variables are assigned to dual links $\widehat{L}_{D-p-1}^{}$ as 
\aln{
Z_{G}^{(p)}(\beta)=\int [d\widehat{U}]e^{-\widehat{S}}\times \left(\prod_{\widehat{L}_{D-p}^{}}\delta_{0,d_{\widehat{L}_{D-p}^{}}^{}\alpha
}^{}
\right)~,
\label{dual form expression}
}
where 
\aln{
d_{\widehat{L}_{D-p}^{}}\alpha=\sum_{\widehat{L}_{D-p-1}^{}}\mathrm{Inc}(\widehat{L}_{D-p}^{},\widehat{L}_{D-p-1}^{})\alpha_{\widehat{L}_{D-p-1}^{}}
}
is the boundary operator and 
\aln{
\widehat{S}^{}=-\sum_{\widehat{L}_{D-p-1}^{}}\log f(\alpha_{\widehat{L}_{D-p-1}^{}}^{})~,\quad \int [d\widehat{U}]=\prod_{\widehat{L}_{D-p-1}^{}}\sum_{\alpha_{\widehat{L}_{D-p-1}^{}}^{}\in \widehat{G}}~.
\label{dual action}
}
In the following, we denote every dual quantity with a hat. 
From the definition of the boundary operator, one can see that the delta functions in Eq.~(\ref{dual form expression}) impose the local flatness conditions for dual link variables. 
Although Eq.~(\ref{dual action}) is not gauge invariant, it can be further rewritten in a gauge invariant way by introducing additional $(D-p-2)$-link variables $\widehat{M}_{\widehat{L}_{D-p-2}^{}}^{}\in \widehat{G}$ as 
\aln{
Z^{(D-p-2)}_{/\widehat{G}}(\widehat{\beta})\coloneq \frac{1}{\mathrm{Vol}[\text{Gauge}]}\int [d\widehat{M}]\int[d\widehat{U}]e^{-\widehat{S}^{(D-p-2)}}\times \prod_{\widehat{\mathrm{P}}_{D-p-1}^{}} \delta_{1,W_{}^{}[\widehat{\mathrm{P}}_{D-p-1}^{}]}^{}~,
\label{dual gauged theory}
}
where 
\aln{
&\widehat{S}_{}^{(D-p-2)}\coloneq 
-\sum_{\mathrm{P}_{D-p-2}^{}}
K(\widehat{M}_{L_{D-p-2}^{}}^{})\log f(\widehat{U}_{\widehat{L}_{D-p-1}^{}}^{})~,\quad 
W[\widehat{\mathrm{P}}_{D-p-1}^{}]=\prod_{\widehat{L}_{D-p-1}^{}\subset \widehat{\mathrm{P}}_{D-p-1}^{}}\widehat{U}_{\widehat{L}_{D-p-1}}^{}~.
}
Here, $/\widehat{G}$ denotes the gauging of $\widehat{G}$, $\widehat{\beta}$ is the dual inverse gauge coupling (determined by $\beta$), and the argument of $f(\alpha)$ is changed from $\alpha$ to $\widehat{U}=e^{i\alpha}\in \widehat{G}$.  
For $G=\mathrm{U}(1)$, we simply identify $\alpha=\widehat{U}\in \mathbb{Z}$. 
Moreover, $K(x)$ is a real function satisfying $K(1)=1$ and 
\aln{
K(\widehat{M}_{L_{D-p-2}^{}}^{'})\log f(\widehat{U}_{\widehat{L}_{D-p-1}^{}}^{'})=K(\widehat{M}_{L_{D-p-2}^{}}^{})\log f(\widehat{U}_{\widehat{L}_{D-p-1}^{}}^{})
\label{gauged gauge invariance}
} 
under the dual gauge transformation 
\aln{
\widehat{M}_{\widehat{L}_{D-p-2}^{}}\quad &\rightarrow \quad \widehat{M}_{\widehat{L}_{D-p-2}^{}}^{'}=g(\widehat{L}_{D-p-2}^{})\widehat{M}_{\widehat{L}_{D-p-2}^{}}~,\quad g(\widehat{L}_{D-p-2}^{})\in \widehat{G}~,
\\
\widehat{U}_{\widehat{L}_{D-p-1}^{}}\quad &\rightarrow \quad \left(\prod_{\widehat{L}_{D-p-2}^{}\subset \partial\widehat{L}_{D-p-1}^{}}g(\widehat{L}_{D-p-2}^{})^{-1}\right)\widehat{U}_{\widehat{L}_{D-p-1}}^{}~. 
}
The explicit functional form of $K(x)$ depends on a choice of $F(\theta)$. 
As long as Eq.~(\ref{gauged gauge invariance}) is satisfied, we can eliminate all $(D-p-2)$-link variables as $\widehat{M}_{D-p-2}^{}=1$ by this gauge transformation and obtain  
\aln{
Z^{(p)}_G(\beta)=Z^{(D-p-2)}_{/\widehat{G}}(\widehat{\beta})~
\label{general KW duality}
} 
up to a normalization factor. 
This is the KW duality between the two lattice higher gauge theories and relates high-temperature phase of one theory to low-temperature phase of the other in general.  
In Appendix~\ref{app:KW}, we discuss several examples. 
\begin{figure}
 \centering
 \includegraphics[scale=0.4]{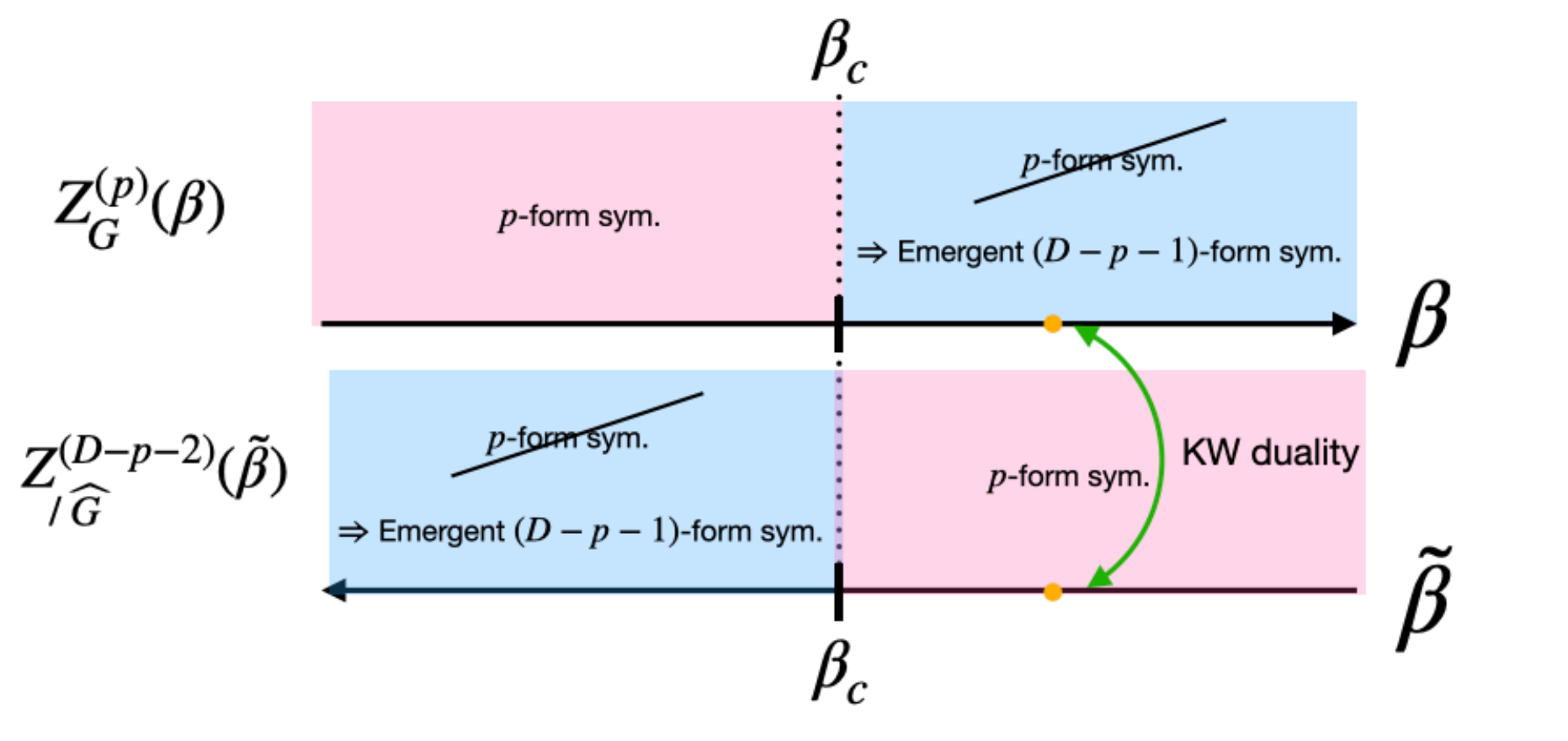}
\caption{
The Kramers-Wannier duality in finite higher gauge theories. 
The pink~(blue) regions correspond to high (low) temperature regions.  
 }
 \label{fig:KW}
\end{figure}

The local flatness condition of the $(D-p-1)$-form gauge field in the dual theory $Z_{/\widehat{G}}^{(D-p-2)}(\widehat{\beta})$ implies that the Wilson-surface operator 
\aln{
W[C_{D-p-1}^{}]=\prod_{\widehat{L}_{D-p-1}^{}\subset \widehat{C}_{D-p-1}^{}}\widehat{U}_{\widehat{L}_{D-p-1}}^{}~
} 
is a topological operator, corresponding to a $p$-form global symmetry $\widehat{\widehat{G}}=G$ as expected from the KW duality~(\ref{general KW duality}). 
In Fig.~\ref{fig:KW}, we show the phase diagram, assuming that there  exists a single phase transition as a function of the gauge coupling.\footnote{
This assumption is nontrivial. 
Many numerical studies support the existence of a Coulomb phase in the intermediate coupling region for $N\geq 5$~\cite{Nguyen:2024ikq}.   
It is quite interesting to discuss such a Coulomb phase in the Landau theory framework.   
}


\subsection{Duality in Landau field theories}
The KW duality~(\ref{general KW duality}) also implies an exact duality between two Landau field theories through the Hubbard-Stratonovich transformation as discussed so far. 
In the continuum limit, this leads to an infrared (IR) duality, that is, two theories describe the same critical phenomena or flow to the same IR critical point, similar to the $2+1$ dimensional particle/vortex duality~\cite{Peskin:1977kp,Karch:2016sxi,Kawana:2024fsn}.         
In the following, we elucidate such an IR duality in Landau field theories for a finite higher-form symmetry.     
To this end, it is necessary to understand the gauging procedure in the Landau theory context.      
For concreteness, we focus on $G=\mathbb{Z}_N^{}$ in the following.   

In order to gauge the $(D-p-2)$-form global symmetry $\widehat{G}$ in $Z^{(D-p-2)}_{\text{Landau}}(\widehat{\beta})$, 
%
we introduce a $(D-p-1)$-form gauge field $A_{D-p-1}^{}$ with a locally flat condition 
\aln{
NA_{D-p-1}^{}=d\Phi_{D-p-2}^{}~,
\label{flat condition}
}
where $\Phi_{D-p-2}^{}$ is a differential $(D-p-2)$-form satisfying $\int_{C_{D-p-1}^{}}d\Phi_{D-p-2}^{}\in 2\pi \mathbb{Z}$, where $C_{D-p-1}^{}$ is a $(D-p-1)$-dimensional closed subspace.  
Equation~(\ref{flat condition}) can be realized by introducing a $(p+1)$-form Lagrangian multiplier $B_{p+1}^{}$ as 
\aln{
ic\int_{\Sigma_D^{}}  (NA_{D-p-1}^{}-d\Phi_{D-p-2}^{})\wedge B_{p+1}^{}~,
\label{gauging p-form}
} 
where $c$ is a real constant that should be determined consistently. 
This action is invariant under the following gauge transformation:
\aln{
A_{D-p-1}^{}\quad \rightarrow \quad A_{D-p-1}^{}+d\Lambda_{D-p-2}^{}~,\quad \Phi_{D-p-2}^{}\quad \rightarrow \quad \Phi_{D-p-2}^{}+N\Lambda_{D-p-2}^{}+d\Lambda_{D-p-3}^{}~,
\label{gauge transformation of BF}
} 
where $\Lambda_{D-p-2}^{}~(\Lambda_{D-p-3}^{})$ is a $(D-p-2)$-~$((D-p-3)$-) form gauge parameter field.  
Equation~(\ref{gauging p-form}) should respect a $\mathbb{Z}_N^{}$ $(D-p-1)$-form global symmetry, whose charged object is the Wilson-surface operator $W[C_{D-p-1}^{}]=\exp\left(i\int_{C_{D-p-1}}A_{D-p-1}^{}\right)$. 
Namely, the coefficient $c$ can be determined by demanding that  Eq.~(\ref{gauging p-form}) is invariant up to $2\pi i n,~n\in \mathbb{Z}$ under the $\mathbb{Z}_N^{}$ $(D-p-1)$-form transformation 
\aln{
A_{D-p-1}^{}\quad \rightarrow \quad A_{D-p-1}^{}+\frac{1}{N}\Lambda_{D-p-1}^{}~,\quad d\Lambda_{D-p-1}^{}=0~,\quad \int_{C_{D-p-1}}\Lambda_{D-p-1}^{}\in 2\pi \mathbb{Z}~.
} 
%
Then, we find $c=(2\pi)^{-1}$ along with the Dirac quantization condition $\int_{C_{p+2}^{}}dB_{p+1}^{}\in 2\pi \mathbb{Z}$, where $C_{p+2}^{}$ is a $(D-q)$-dimensional closed subspace.  
Besides, the variation with respect to $\Phi_{D-p-2}^{}$ leads to the condition $dB_{p+1}^{}=0$, i.e., $B_{p+1}=dB_{p}^{}$ locally, and we obtain 
\aln{
i\frac{N}{2\pi}\int_{\Sigma_D^{}} A_{D-p-1}^{}\wedge dB_{p}^{}~, 
}
which is nothing but a $\mathrm{BF}$-type topological field theory, as expected.  

Next, let us consider the matter part in the Landau field theory $Z^{(D-p-2)}_{\text{Landau}}(\widehat{\beta})$. 
%
%
The key observation is that the kinetic term is expressed in terms of the area derivative~(\ref{projection}), which is a $(D-p-1)$-form tensor in spacetime.  
Thus, we can naturally find a covariant kinetic term as 
\aln{
{\cal N}\int {\cal D}X\frac{1}{\beta \mathrm{Vol}[C_{D-p-2}^{}]}\int_{S^{D-p-2}}^{} d^{D-p-2}\xi\sqrt{h(\xi)}~\phi[C_{D-p-2}^{}]^*\frac{\delta^2}{\delta D^{\mu}(\xi)^*\delta D_{\mu}^{}(\xi)}\phi[C_{D-p-2}^{}]~,
}
where the covariant derivative $D_\mu^{}$ is defined by 
\aln{
D_\mu^{}\phi[C_{D-p-2}^{}]\coloneq \{X^{\nu_1^{}},\cdots,X^{\nu_{D-p-2}}\}\left(\frac{\delta}{\delta \sigma^{\mu\nu_1^{}\cdots \nu_{D-p-2}^{}}(\xi)}-iA_{\mu \nu_1^{}\cdots \nu_{D-p-2}^{}}(X(\xi))\right)\phi[C_{D-p-2}^{}]~.
\label{def covariant derivative}
}
One can check that this transforms as
\aln{
D_\mu^{}\phi[C_{D-p-2}^{}]\quad \rightarrow \quad e^{i\int_{C_{D-p-2}^{}}\Lambda_{D-p-2}^{}}D_\mu^{}\phi[C_{D-p-2}^{}]
} 
under the gauge transformation
\aln{
\phi[C_{D-p-2}^{}]\quad \rightarrow \quad e^{i\int_{C_{D-p-2}^{}}\Lambda_{D-p-2}^{}}\phi[C_{D-p-2}^{}]~
\label{phi gauge transformation}
}
along with Eq.~(\ref{gauge transformation of BF}). 
Besides, the potential $U(\phi,\phi^*)$ can be coupled to the gauge field  in a gauge invariant manner as 
\aln{
U(\phi[C_{D-p-2}^{}],\phi[C_{D-p-2}^{}]^*)\quad \rightarrow \quad U\left(e^{i\int_{M_{D-p-1}}A_{D-p-1}^{}}\phi[C_{D-p-2}^{}],e^{-i\int_{M_{D-p-1}}A_{D-p-1}^{}}\phi[C_{D-p-2}^{}]^*\right)~,
} 
where $M_{D-p-1}^{}$ is the $(D-p-1)$-dimensional minimal surface bounded by $C_{D-p-2}^{}$.  
One can easily check that the combination $e^{i\int_{M_{D-p-1}}A_{D-p-1}^{}}\phi[C_{D-p-2}^{}]$ is gauge invariant. 

To summarize, the gauged version of $Z^{(D-p-2)}_{\text{Landau}}(\widehat{\beta})$ is  
\aln{
Z_{\text{Landau}/\mathbb{Z}_N^{}}^{(D-p-2)}(\widehat{\beta})\coloneq \int {\cal D}\phi {\cal D}\phi^*\int {\cal D}A_{D-p-1}^{} {\cal D}B_{p}^{}~e^{-S_{\rm gauged}^{(D-p-2)}}~,
\label{partition function of gauged Landau}
}
where 
\aln{
S_{\rm gauged}^{(D-p-2)}=&-i\frac{N}{2\pi}\int_{\Sigma_D^{}}A_{D-p-1}^{}\wedge dB_{p}^{}
\nn
&+{\cal N}\int {\cal D}X\bigg[U\left(e^{i\int_{M_{D-p-1}}A_{D-p-1}^{}}\phi[C_{D-p-2}^{}],e^{-i\int_{M_{D-p-1}}A_{D-p-1}^{}}\phi[C_{D-p-2}^{}]^*\right)
\nn
&-\frac{1}{\beta \mathrm{Vol}[C_{D-p-2}^{}]}\int_{S^{D-p-2}}^{} d^{D-p-2}\xi\sqrt{h(\xi)}~\phi[C_{D-p-2}^{}]^*\frac{\delta^2}{\delta D^{\mu}(\xi)\delta D_{\mu}^{}(\xi)}\phi[C_{D-p-2}^{}]\bigg]~.
\label{Landau theory of gauged theory}
}
Although we have derived Eq.~(\ref{partition function of gauged Landau}) heuristically above, it can be directly derived from Eq.~(\ref{dual gauged theory}) by performing the Hubbard-Stratonovich transformation and taking the continuum limit. 
This is shown in Appendix~\ref{app:gauged}.

Finally, the KW duality~(\ref{general KW duality}) suggests that the corresponding Landau field theories is IR dual as 
\aln{
Z_{\text{Landau}}^{(p)}(\beta)\sim Z_{\text{Landau}/\mathbb{Z}_N^{}}^{(D-p-2)}(\widehat{\beta})~,
\label{duality of Landau theories}
}
where $\sim$ means that this is valid only near the critical point $\beta\sim \widehat{\beta}\sim \beta_c^{}$.  
Both theories have a $p$-form global symmetry, and the correspondence of charged and symmetry operators are given as 
\aln{
\begin{dcases}
\phi[C_p^{}]\quad \leftrightarrow \quad e^{i\int_{C_{p}^{}}B_{p}^{}}
& (\text{charged operators})
\\
U[C_{D-p-1}^{}]\quad \leftrightarrow \quad e^{i\int_{C_{D-p-1}^{}}A_{D-p-1}^{}} & (\text{symmetry operators})
\end{dcases}
}   
Since the duality~(\ref{duality of Landau theories}) involves closed objects with different dimensions, this can be interpreted as an extension of the particle/vortex duality for $\mathrm{U}(1)$ higher-form global symmetry~\cite{Peskin:1977kp,Kawana:2024fsn}.   
In particular, one can see that field theory of closed $p$-branes ($p\geq 1$) is can be dual to field theory of particles when 
\aln{
D-p-2=0\quad \therefore \quad p=D-2~.
}
For example, when $D=3\leftrightarrow p=1$, $Z^{(1)}_{\text{Landau}}(\beta)$ is a closed string field theory with a $\mathbb{Z}_N^{}$ $1$-form global symmetry, while the dual theory $Z_{\text{Landau}/\mathbb{Z}_N^{}}^{(0)}(\widehat{\beta})$ is a Higgs theory coupled to the $\mathbb{Z}_N^{}$ gauge theory.    
This indicates that the confinement/deconfinement transition of the $3$-dimensional $\mathbb{Z}_N^{}$ lattice gauge theory is actually a particle-like transition rather than a string-like one, as already confirmed by numerical studies~\cite{Hasenbusch:1992eu}.     
\begin{table}[t]
\vspace{-1cm}
\begin{adjustbox}{center}
\begin{tabular}{|c||c|c|c|c|}\hline
      & Order Parameter  & Massive modes         
            \\\hline\hline
$Z^{(p)}_{\text{Ising}}(\beta)$ (unbroken phase) & Area law of $\phi[C_{p}^{}]$    &   $\phi[C_{p}^{}]$ excitations  
 \\\hline
$Z^{(D-p-2)}_{\text{Ising}/\mathbb{Z}_N^{}}(\widehat{\beta})$ (broken phase) &  Perimeter law of $e^{i\int_{C_p^{}}B_p^{}}$  &  $p$-dim static topological solitons      
 \\\hline
 \end{tabular}
  \end{adjustbox}
\caption{
IR duality in the unbroken phase in $Z^{(p)}_{\text{Ising}}(\beta)$. 
This corresponds to the unbroken phase in $Z^{(D-p-2)}_{\text{Ising}/\mathbb{Z}_N^{}}(\widehat{\beta})$.  
}
\label{tab:confined}
\end{table}
\begin{table}[!t]
\begin{adjustbox}{center}
\begin{tabular}{|c||c|c|c|c|}\hline
      &  Order Parameter & Massive modes          
            \\\hline\hline 
$Z^{(p)}_{\text{Ising}}(\beta)$ (broken phase) &  Perimeter law of $\phi[C_{p}^{}]$   &   $(D-p-2)$-dim static topological solitons  
 \\\hline
$Z^{(D-p-2)}_{\text{Ising}/\mathbb{Z}_N^{}}(\widehat{\beta})$ (unbroken phase)  & Area law of  $e^{i\int_{C_p^{}}B_p^{}}$  &  $\phi[C_{D-p-3}^{}]$ excitations    
 \\\hline
 \end{tabular}
 \end{adjustbox}
\caption{
IR duality in the broken phase in $Z^{(p)}_{\text{Ising}}(\beta)$. 
This corresponds to the broken phase in $Z^{(D-p-2)}_{\text{Ising}/\mathbb{Z}_N^{}}(\widehat{\beta})$.  
}
\label{tab:deconfined}
\end{table}

In Tables~\ref{tab:confined} and \ref{tab:deconfined}, we summarize the IR duality. 
Although we have not fully investigated quantum phases of these theories, the correspondence of symmetry, order parameters, and excitation modes provides compelling evidence to support this IR duality from universality point of view.  
More dedicated studies are left for future investigations.
%
%

\section{Summary}\label{sec6}

 We have derived the continuum Landau field theory for the lattice higher gauge theory and studied the phases of the gauge theory at the mean-field level. 
 By solving the functional equation of motion, we have shown that the classical solution exhibits the area (perimeter) law in the strong (weak) gauge coupling limit. 
 Besides, we have presented explicit constructions of topological defects,  which are higher-dimensional analogues of vortex/domain-wall solutions in ordinary field theories with $0$-form global symmetries.  
 The existence of topological defects has a critical impact on the phase structure in higher gauge theories, and we have derived the Coleman-Mermin-Wagner theorem for higher-form global symmetries in the Landau theory framework.  

Moreover, we have formulated the Kramers-Wannier duality in our lattice setup and elucidated how it can lead IR duality in the corresponding  Landau field theories.  
 This IR duality relates a field theory of $p$-dimensional closed objects with $p$-form  global symmetry to a field theory of $(D-p-2)$-dimensional closed objects with a gauged higher-form symmetry, and can be viewed as a discrete-symmetry analog of the particle-vortex duality~\cite{Peskin:1977kp,Kawana:2024fsn}.  
Although a complete understanding of quantum theory of extended objects remains elusive, these results suggest that higher-form symmetries plays a crucial role for quantum many-body physics of extended objects.

\section*{Acknowledgements}
We thank Yoshimasa Hidaka 
for fruitful discussions and comments. 
This work is supported by KIAS Individual Grants, Grant No. 090901.   
%

\appendix 
\section{Higher gauge theories}\label{app:operator relation}
In this Appendix, we check the operator relation~(\ref{operator relation}) in the continuum higher gauge theories. 
In this Appendix, we employ the Lorentzian signature $(-,+,\cdots,+)$. 

The $\mathrm{U}(1)$ $p$-form Maxwell theory is 
\aln{
S_p^{}[A_p^{}]=-\frac{1}{2g^2}\int_{\Sigma_D^{}}F_{p+1}^{}\wedge \star F_{p+1}^{}~,\quad F_{p+1}^{}=A_p^{}~. 
}
In the path-integral formalism, the VEV of the Wilson-surface operator in the presence of symmetry operator is
\aln{
\langle U_\theta^{}[C_{D-p-1}^{}]W[C_p^{}]\rangle &\coloneq \frac{1}{Z}\int {\cal D}A_{p}^{}e^{iS_p^{}[A_p^{}]}U_\theta^{}[C_{D-p-1}^{}]W[C_p^{}]
\\
&=\frac{1}{Z}\int {\cal D}A_{p}^{}e^{iS_p^{'}[A_p^{}]}W[C_p^{}]~,
}
where
\aln{
S_p^{'}[A_p^{}]&=S_p^{}[A_p^{}]+\frac{\theta}{g^2}\int_{\Sigma_D^{}}\delta_{p+1}^{}(C_{D-p-1}^{})\wedge \star F_{p+1}^{}
\nn
&=-\frac{1}{2g^2}\int_{\Sigma_D^{}}F'_{p+1}\wedge \star F'_{p+1}~,\quad F'_{p+1}=F_{p+1}^{}-\theta \delta_{p+1}^{}(C_{D-p-1}^{})~.
}
Here, $\delta_{p+1}^{}(C_{D-p-1}^{})$ is the Poincare dual form of $C_{D-p-1}^{}$. 
Denoting a $(D-p)$-dimensional open manifold bounded by $C_{D-p-1}^{}$ as $M_{D-p}^{}$, we have  
\aln{
\delta_{p+1}^{}(C_{D-p-1}^{})=d\delta_{p}^{}(M_{D-p}^{})~,
} 
which leads to $F'_{p+1} =dA'_p$ with $A'_p=A_p^{}+\theta \delta_{p}^{}(M_{D-p}^{})$.
Then, changing the path-integral variable as $A'_p\rightarrow A_p^{}$, we obtain
\aln{\langle U_\theta^{}[C_{D-p-1}^{}]W[C_p^{}]\rangle=\frac{1}{Z}\int {\cal D}A_{p}^{}e^{iS_p^{}[A_p^{}]}W[C_p^{}]\exp\left(i\theta \int_{C_p^{}}\delta_p^{}(M_{D-p}^{})\right)=e^{i\theta {\rm Link}[C_p^{},C_{D-p-1}^{}]}\langle W[C_p^{}]\rangle~,
}
which proves Eq.~(\ref{operator relation}) in the path-integral formalism. 

Next, the $\mathbb{Z}_N^{}$ $p$-form gauge theory is
\aln{
S_{\rm BF}^{}[A_p^{},B_{D-p-1}^{}]=\frac{N}{2\pi}\int_{\Sigma_D^{}} dA_{p}^{}\wedge B_{D-p-1}^{}~.
}
In this case, the symmetry operator is 
\aln{
U_m^{}[C_{D-p-1}^{}]=e^{i\frac{2\pi m}{N}\int_{C_{D-p-1}^{}}B_{D-p-1}^{}}~,\quad m\in \mathbb{Z}~.
}
Repeating the same calculation as above, we have
\aln{
S'_{\rm BF}[A_p^{},B_{D-p-1}^{}]&=S_{\rm BF}^{}[A_p^{},B_{D-p-1}^{}]+\frac{2\pi m}{N}\int_{\Sigma_D^{}} \delta_{p+1}^{}(C_{D-p-1}^{})\wedge B_{D-p-1}^{} 
\nn
&=S_{\rm BF}^{}\left[A_p^{}+\frac{2\pi m}{N}\delta_p^{}(M_{D-p}^{}),B_{D-p-1}^{}\right]~,
}
\aln{
\therefore\quad \langle U_\theta^{}[C_{D-p-1}^{}]W[C_p^{}]\rangle=e^{i\frac{2\pi m}{N}{\rm Link}[C_p^{},C_{D-p-1}^{}]}\langle W[C_p^{}]\rangle~,
}
which is a $\mathbb{Z}_N^{}$ transformation of the Wilson surface operator.  

\section{Effective action in the broken phase}\label{app:effective theory}

Here we present the detailed calculation of the effective action of the phase modulation $A_p^{}$. 
In the following, we consider the flat space $\Sigma_D^{}=\mathbb{R}^D$ for simplicity. 
By putting Eq.~(\ref{phase modulation}) into the kinetic term in the continuum action~(\ref{continuum action}), we have 
\aln{
\frac{v^2}{2\beta}{\cal N}\int {\cal D}X\frac{1}{\mathrm{Vol}[C_p^{}](p!)^2}\int_{S^p}d^p\xi  \sqrt{h(\xi)} \{F_{\mu\nu_1^{}\cdots \nu_p^{}}^{}(X(\xi))E^{\nu_1^{}\cdots \nu_p^{}}\}\{{F^\mu}_{\rho_1^{}\cdots \rho_p^{}}^{}(X(\xi))E^{\rho_1^{}\cdots \rho_p^{}}\}~,
\label{kinetic term of A}
} 
where $F_{p+1}^{}=dA_p^{}=\frac{1}{(p+1)!}F_{\mu_1^{}\cdots \mu_{p+1}}^{}dX^{\mu_1^{}}\wedge\cdots \wedge dX^{\mu_{p+1}^{}}$.   
Introducing the center-of-mass coordinates and relative coordinates as 
\aln{
x^\mu &\coloneqq\frac{1}{\mathrm{Vol}[C_p^{}]}\int_{S^p} d^p\xi\sqrt{h}X^\mu(\xi)~,  \quad Y^\mu(\xi)\coloneqq X^\mu(\xi)-x^\mu~,
}
we have 
\aln{
F_{\mu\nu_1^{}\cdots \nu_p^{}}^{}(X(\xi)){F^\mu}_{\rho_1^{}\cdots \rho_p^{}}^{}(X(\xi))=\int \frac{d^Dk}{(2\pi)^D}e^{ik\cdot (x+Y)}
{\tilde{F}}_{\mu\nu_1^{}\cdots \nu_p^{}}^{}(k){{\tilde{F}^\mu}}_{\rho_1^{}\cdots \rho_p^{}}(k)~,
}
where $\tilde{F}_{\mu_1^{}\cdots \mu_{p+1}^{}}^{}(k)$ is the Fourier mode of $F_{\mu_1^{}\cdots \mu_{p+1}^{}}^{}(x)$.  
By putting this into Eq.~(\ref{kinetic term of A}), we obtain
\aln{
\int d^Dk\left(\int \frac{d^Dx}{(2\pi)^D}~e^{ik\cdot x}\right)
&{\tilde{F}}_{\mu\nu_1^{}\cdots \nu_p^{}}^{}(k){{\tilde{F}^\mu}}_{\rho_1^{}\cdots \rho_p^{}}(k)
\nn
\times &{\cal N}\int {\cal D}Y\frac{1}{\mathrm{Vol}[C_p^{}]}\int_{S^p}d^p\xi \sqrt{h}
e^{ik\cdot Y(\xi)}E^{\nu_1^{}\cdots \nu_p^{}}(Y(\xi))E^{\rho_1^{}\cdots \rho_p^{}}(Y(\xi))
\nn
={\tilde{F}}_{\mu\nu_1^{}\cdots \nu_p^{}}^{}(0){{\tilde{F}^\mu}}_{\rho_1^{}\cdots \rho_p^{}}(0)
& \frac{{\cal N}}{\mathrm{Vol}[C_p^{}]}\int_{S^p}d^p\xi \sqrt{h}\langle E^{\nu_1^{}\cdots \nu_p^{}}E^{\rho_1^{}\cdots \rho_p^{}}\rangle~.
}
Note that $E^{\nu_1^{}\cdots \nu_p^{}}(Y(\xi))$ depends only on the relative coordinates $\{Y^{\mu}(\xi)\}_{\mu=1}^D$. 
Assuming that spacetime symmetry is not spontaneously broken, the expectation value can be expanded as a function of momentums $\{k_\mu^{}\}_{\mu=1}^D$ as    
\aln{
\langle E^{\nu_1^{}\cdots \nu_p^{}}E^{\rho_1^{}\cdots \rho_p^{}}\rangle &=\frac{1}{{(p!)^3}}\sum_{\sigma,\sigma'\in S_p^{}}\mathrm{sgn}(\sigma)\mathrm{sgn}(\sigma') \bigg[c_0^{}(k^2)\eta^{\nu_{\sigma(1)}^{}\rho_{\sigma'(1)}^{}}\cdots \eta^{\nu_{\sigma(p)}^{}\rho_{\sigma'(p)}^{}}
\nn
&\qquad+c_1^{} (k^2)k^{\nu_{\sigma(1)}^{}}k^{\rho_{\sigma'(1)}^{}}\eta^{\nu_{\sigma(2)}^{}\rho_{\sigma'(2)}^{}}\cdots \eta^{\nu_{\sigma(p)}^{}\rho_{\sigma'(p)}^{}}+\cdots\bigg]~,
\label{k expansion}
}
where $c_0^{}(k^2)$ and $c_1^{}(k^2)$ are functions of $k^2$.  
In the low-energy limit, we can further neglect the $k$ dependence in these functions, and the leading action is  
\aln{
{\cal N}c_0^{}(0)\frac{v^2}{2\beta(p!)^3}{\tilde{F}}_{\mu\nu_1^{}\cdots \nu_p^{}}^{}(0){{\tilde{F}^\mu}}_{\rho_1^{}\cdots \rho_p^{}}(0)\eta^{\nu_{\sigma(1)}^{}\rho_{\sigma'(1)}^{}}\cdots \eta^{\nu_{\sigma(p)}^{}\rho_{\sigma'(p)}^{}}
={\cal N}c_0^{}(0)\frac{v^2}{2\beta}\int_{\Sigma_D^{}}F_{p+1}^{}\wedge \star F_{p+1}^{}~, 
}
which corresponds to the $p$-form Maxwell theory~(\ref{naive effective action}) by choosing the normalization factor as ${\cal N}c_0^{}(0)=1$.

\section{Examples of Kramers-Wannier duality}\label{app:KW}

\noindent {\bf $2$-dimensional $\mathbb{Z}_2^{}$ Ising model}\\
The $2$-dimensional $\mathbb{Z}_2^{}$ Ising model corresponds to $D=2$, $p=0$, $G=\mathbb{Z}_2^{}$, and 
\aln{
F(\theta)=\begin{cases} -1 & \text{for } \theta\equiv 0~(\text{mod $2$})
\\
1 & \text{for } \theta\equiv 1~(\text{mod $2$})
\end{cases}~.
\label{Z2 model}
}
In this case, the Fourier mode is 
\aln{
f(\alpha)=\frac{1}{2}\sum_{\theta=0,1}\chi_\alpha^{}(\theta)e^{-\beta F(\theta)}=\begin{cases} \cosh (\beta )& \text{for }\alpha\equiv  0~(\text{mod $2$})
\\
\sinh(\beta) & \text{for }\alpha\equiv 1~(\text{mod $2$})
\end{cases},
}
or
\aln{
\sqrt{\frac{2}{\sinh(2\beta)}}f(\alpha)=\begin{cases}(\tanh(\beta))^{-\frac{1}{2}}& \text{for }\alpha\equiv 0~(\text{mod $2$})
\\
(\tanh(\beta))^{\frac{1}{2}} & \text{for }\alpha\equiv 1~(\text{mod $2$})
\end{cases},
}
by which we can identify the dual coupling as
\aln{(\coth(\beta))^{\frac{1}{2}}=e^{\widehat{\beta}}\quad \leftrightarrow \quad \sinh(2\beta)\sinh(2\widehat{\beta})=1~.
\label{relation of coupling}
} 
A critical point is determined by $\beta=\widehat{\beta}=\beta_c^{}$, i.e.,
\aln{
\sinh(2\beta_c^{})=1\quad \therefore \quad \beta_c^{}=\frac{1}{2}\log(1+\sqrt{2})=0.440~.
\label{critical coupling}
}

\

\noindent {\bf $4$-dimensional $\mathbb{Z}_2^{}$ pure gauge theory}\\
The $4$-dimensional $\mathbb{Z}_2^{}$ pure lattice gauge theory corresponds to $D=4$, $p=1$, $G=\mathbb{Z}_2^{}$, and $F(\theta)$ is the same as Eq.~(\ref{Z2 model}). 
This system is also self dual, and we obtain the same coupling relation as Eq.~(\ref{relation of coupling}) and the critical coupling~(\ref{critical coupling}), assuming that there is a single phase transition. 
This applies to any $\mathbb{Z}_2^{}$ self-dual model.   
 
 \
 
 \noindent {\bf $3$-dimensional pure compact $\mathrm{U}(1)$ gauge theory}\\
 The $3$-dimensional pure compact $\mathrm{U}(1)$ lattice gauge theory corresponds to $D=3$, $p=1$, $G=\mathrm{U}(1)$, and $F(\theta)=1-\cos(\theta),~\theta\in [-\pi,\pi]$. 
In the low temperature limit $\beta\rightarrow \infty$, we can use the Villain formula as
\aln{
&\exp\left(-\beta\sum_{\mathrm{P}_1^{}} \left(1-\cos(d\theta )\right)\right)\approx \left(\prod_{\mathrm{P}_1^{}}\sum_{n_{\mathrm{P}_1^{}}\in \mathbb{Z}}\right)\exp\left(-\sum_{\mathrm{P}_1^{}}(d\theta-n_{\mathrm{P}_1^{}}^{})^2\right)
\nn
=&\left(\prod_{\mathrm{P}_1^{}}\sqrt{\frac{2\pi}{\beta}}\sum_{m_{\mathrm{P}_1^{}}^{}\in \mathbb{Z}}\right)\exp\left(-\sum_{\mathrm{P}_1^{}}\left[\frac{2\pi^2}{\beta}m_{\mathrm{P}_1^{}}^2-2\pi im_{\mathrm{P}_1^{}}d\theta\right]\right)~.
}
where $d\theta=\sum_{L_1^{}\subset \mathrm{P}_1^{}}^{}\mathrm{sgn}(L_1^{})\theta_{L_1^{}}$ is the discretized exterior derivative, and we have used the mathematical formula 
\aln{
\sum_{n\in \mathbb{Z}}e^{-a(y-n)^2}=\sum_{k\in \mathbb{Z}}\sqrt{\frac{\pi}{a}}e^{-\frac{\pi^2}{a}k^2-2\pi iky}~.
}
in the second line, which can be derived from the Poisson resummation formula. 
Note that $m_{\mathrm{P}_1^{}}^{}$ is a $\mathbb{Z}$-valued $2$-form field defined on $1$-plaquettes.    
The integration of $\theta_{L_1^{}}$ leads to 
\aln{
d\star m_{\mathrm{P}_1^{}}=0\quad \therefore \quad \star m_{\mathrm{P}_1^{}}=d\Phi_{\widehat{i}}^{}~,
}
where $\Phi_{\widehat{i}}^{}$ is a $\mathbb{Z}$-valued scalar defined on dual sites $\widehat{i}\in \widehat{\Lambda}_3^{}$ .  
Now, the partition function is written as 
\aln{
&Z_{\text{Ising}}^{(1)}(\beta)\approx \left(\prod_{\widehat{i}\in \widehat{\Lambda}_3^{}}\sqrt{\frac{2\pi}{\beta}}\sum_{\Phi_{\widehat{i}} \in \mathbb{Z}}\right)\exp\left(-\frac{2\pi^2}{\beta}\sum_{\widehat{j}\in \widehat{\Lambda}_3^{}}d\Phi_{\widehat{j}}^{}\wedge \star d\Phi_{\widehat{j}}^{}
\right)
\nn
=&\left(\prod_{\widehat{i}\in \widehat{\Lambda}_3^{}}\sqrt{\frac{2\pi}{\beta}}\int_{-\infty}^{\infty}d\phi_{\widehat{i}}^{}\sum_{n_{\widehat{i}}^{}\in \mathbb{Z}}\right)\exp\left(-\frac{2\pi^2}{\beta}\sum_{\widehat{i}\in \widehat{\Lambda}_3^{}}d\phi_{\widehat{i}}^{} \wedge \star d\phi_{\widehat{i}}^{}+2\pi i \sum_{\widehat{i}\in \widehat{\Lambda}_3^{}}n_{\widehat{i}}^{} \phi_{\widehat{i}}^{}
\right)~,
\label{scalar with monopoles}
}
where we have used 
\aln{
\sum_{n\in \mathbb{Z}}g(n)=\sum_{k\in \mathbb{Z}}\int_{-\infty}^{\infty}dx~g(x)e^{-2\pi ixk}~,
} 
which can be also derived from the Poisson resummation formula. 
The Lagrangian can be further rewritten as 
\aln{
\frac{2\pi^2}{\beta}\phi_{\widehat{i}} \Box \phi_{\widehat{i}}^{}+2\pi i n_{\widehat{i}}^{} \phi_{\widehat{i}}^{}
=\frac{2\pi^2}{\beta}\left(\phi+\frac{1}{2\pi} i\beta _{\widehat{i}}^{}\Box^{-1}\right)\Box \left(\phi_{\widehat{i}}^{}+\frac{1}{2\pi} i\beta \Box^{-1}n_{\widehat{i}}^{}\right)+\frac{\beta}{2}n_{\widehat{i}}^{}\Box^{-1}n_{\widehat{n}}^{}~,
} 
where $\Box$ is the $3$-dimensional d'Alembert operator. 
The integration of $\{\phi_{\widehat{i}}^{}\}$ leads to  
\aln{
Z(\beta)\approx \left(\prod_{\widehat{i}\in \widehat{\Lambda}_3^{}}\sum_{n_{\widehat{i}}^{}\in \mathbb{Z}}\right)\exp\left(\frac{\beta}{2} \sum_{\widehat{i}\in \widehat{\Lambda}_3^{}} n_{\widehat{i}}^{}\Box^{-1}n_{\widehat{i}}^{}\right)~.
}  
In the continuum limit, the summation of monopole gas can be also written as
\aln{
\left(\prod_{\widehat{i}\in \widehat{\Lambda}_3^{}}\sum_{n_{\widehat{i}}^{}\in \mathbb{Z}}\right)=\sum_{N=0}^\infty \frac{1}{N!}\prod_{i=1}^N\left(\int d^3x_i^{}\sum_{q_i^{}\in \mathbb{Z}}\right) 
} 
with $n_{\widehat{i}}^{} \rightarrow n(x)=\sum_{i=1}^{N}q_i^{}\delta^{(3)}(x-x_i^{})$.
In the $3$-dimensional flat Euclidean space, the propagator is $\Box^{-1}=-\frac{1}{4\pi |x|}$, and we obtain
\aln{
Z^{(1)}_{\text{Ising}}(\beta)\approx \sum_{N=0}^{\infty}\frac{e^{\beta \mu N}}{N!}\left(\prod_{i=1}^N\int d^3x_i^{}\sum_{q_i^{}\in \mathbb{Z}}\right)\exp\left(-\frac{\pi\beta}{4}\sum_{i<j}\frac{q_i^{}q_j^{}}{|x_i^{}-x_j^{}|}
\right)~,
}
where the self-energy was renormalized into the chemical potential $\mu$.  
This represents the partition function of monopole gas. 
Namely, the compact $\mathrm{U}(1)$ gauge theory is dual to the monopole-gas system, and it is equivalent to a massive scalar theory as discussed in subsection~\ref{sec:CMW}.

\section{Landau theory for gauged higher gauge theory}\label{app:gauged}

We derive the Landau theory for the gauged higher gauge theory~(\ref{dual gauged theory}).  
The flow of the derivation is parallel to the ungauged case by extending the definition of the translation operator~(\ref{translation operator}). 
In the following, we simply denote $(D-p-1)$ as $q$ and take the Wilson action~(\ref{lattice Ising model}) as {\it ungauged} $q$-form higher gauge theory. 
Besides, the minimal $(q+1)$-dimensional surface bounded by a $q$-dimensional closed surface $\widehat{C}_q^{}$ is expressed by $\widehat{D}_{q+1}^{}$. 
Although every dual quantity is denoted with a hat below, one can also omit it by regarding the dual lattice $\widehat{\Lambda}_D^{}$ as an original lattice. 

For a given $q$-dimensional surface $\widehat{C}_q^{}$, we define a Wilson open-surface operator of the $(q+1)$-form gauge field by 
\aln{
\tilde{W}[\widehat{D}_{q+1}^{}]\coloneq \prod_{\widehat{L}_{q+1}^{}\in \widehat{M}_{q+1}^{}}\widehat{M}_{\widehat{L}_{q+1}^{}}~,\quad \widehat{M}_{\widehat{L}_{q+1}^{}}\in \widehat{G}~.
}
Then, we can extend the definition of the translation operator $\Pi_{\widehat{\mathrm{P}}_q^{}}$ as 
\aln{
\Pi_{\widehat{\mathrm{P}}_q^{}}^{}\tilde{W}[\widehat{D}_{q+1}^{}]W[\widehat{C}_q^{}]\coloneq \begin{cases}\tilde{W}[\widehat{D}_{q+1}^{}+\widehat{L}_{q+1}^{}]W[\widehat{C}_q^{}+\widehat{\mathrm{P}}_q^{}] & \text{for } \widehat{C}_q^{}+\widehat{\mathrm{P}}_q^{}\in \widehat{\Gamma}_q^{}
\\
0 & \text{for } \widehat{C}_q^{}+\widehat{\mathrm{P}}_q^{}\notin \widehat{\Gamma}_p^{}
\end{cases}~,
}
where $\widehat{L}_{q+1}^{}$ is the dual $(q+1)$-link satisfying $\partial \widehat{L}_{q+1}^{}=\widehat{\mathrm{P}}_q^{}$.  
By repeating the same calculation as Eq.~(\ref{proof}), we can express the gauged lattice action in a quadratic form as    
\aln{
\widehat{S}^{(q)}=\text{constant}-\frac{\beta}{2}\sum_{\widehat{C}_q^{}\in \widehat{\Gamma}_q^{}}w[\widehat{C}_q^{}](\tilde{W}[\widehat{D}_{q+1}^{}]W[\widehat{C}_q^{}])^* (k_0^{}+\hat{H})w[\widehat{C}_q^{}]^*(\tilde{W}[\widehat{D}_{q+1}^{}]W[\widehat{C}_q^{}])~.
}
Performing the HS transformation, we obtain 
\aln{
Z_{/\widehat{G}}^{(q)}=\int [d\widehat{M}]\int [d\phi]&\exp\left\{-\sum_{\widehat{C}_q^{}\in \widehat{\Gamma}_q^{}}\left(\frac{2}{\beta}\phi[\widehat{C}_q^{}]^* (k_0^{}+\hat{H})^{-1}\phi[\widehat{C}_q^{}]+V(\tilde{W}^*\phi,\tilde{W}\phi^*)\right)\right\}\times \prod_{\text{all}~\widehat{\mathrm{P}}_{q+1}^{}} \delta_{\tilde{W}_{}^{}[\widehat{\mathrm{P}}_{q+1}^{}],1}^{}~,
\label{HS in gauged theory}
}
where 
\aln{
V(\tilde{W}^*\phi,\tilde{W}\phi^*)=-\log\bigg[\frac{1}{\mathrm{Vol}[\text{Gauge}]}&\int [d\widehat{U}]\exp\left(-\sum_{\widehat{C}_q^{}\in \widehat{\Gamma}_q^{}}w[\widehat{C}_q^{}](\tilde{W}[\widehat{D}_{q+1}^{}]W[\widehat{C}_q^{}])^*\phi[\widehat{C}_q^{}]+{\rm h.c.}\right)
\bigg]~. 
}
In particular, the covariant generalization of the area derivative~(\ref{def of area derivative}) is 
\aln{
\frac{\delta \psi[\widehat{C}_q^{}]}{\delta \Sigma^{\mu_1^{}\cdots \mu_{p+1}}(\hat{i})}\coloneqq \frac{\tilde{W}[\widehat{L}_{\mu_1^{}\cdots \mu_{q+1}}^{}(\hat{i})]\psi[\widehat{C}_q^{}+\widehat{\mathrm{P}}_{\mu_1^{}\cdots \mu_{q+1}}^{}(\hat{i})]-\psi[\widehat{C}_q^{}]}{a^{q+1}}~,
\label{}
}
where $\widehat{L}_{\mu_1^{}\cdots \mu_{q+1}}^{}(\hat{i})$ is the dual $(p+1)$-link satisfying $\partial \widehat{L}_{\mu_1^{}\cdots \mu_{q+1}}^{}(\hat{i})=\widehat{\mathrm{P}}_{\mu_1^{}\cdots \mu_{q+1}}^{}(\hat{i})$. 
In the continuum-limit, this becomes
\aln{
=&\frac{e^{-i a^{q+1} (A_{q+1}^{})_{\mu_1^{}\cdots \mu_{q+1}}^{}}\psi[\widehat{C}_q^{}+\widehat{\mathrm{P}}_{\mu_1^{}\cdots \mu_{q+1}}^{}(\hat{i})]-\psi[\widehat{C}_q^{}]}{a^{p+1}}
\nn
=&\frac{\delta \psi[\widehat{C}_q^{}]}{\delta \sigma^{\mu_1^{}\cdots \mu_{q+1}^{}}(\xi)}-i(A_{q+1}^{})_{\mu_1^{}\cdots \mu_{q+1}}^{}\psi[\widehat{C}_q^{}]+{\cal O}(a^{q+1})~,
}  
which coincides with Eq.~(\ref{def covariant derivative}). 
This proves that the continuum Landau theory of Eq.~(\ref{HS in gauged theory}) is Eq.~(\ref{Landau theory of gauged theory}).  

\bibliographystyle{TitleAndArxiv}
\bibliography{Bibliography}

\end{document}